# Predicting Organic Solar Cell Performance and Stability from Fast, Morphology-aware Current-Voltage Modeling

Yasin Ameslon[1,2], Larry Lüer[3], Jens Harting[1,2,4], Olga Wodo[5], Olivier J.J. Ronsin[1]

[1]Helmholtz-Institute Erlangen-Nürnberg for Renewable Energy (IET-2), Forschungszentrum Jülich, Erlangen, Germany

[2]Department of Chemical and Biological Engineering, Friedrich-Alexander-Universität Erlangen-Nürnberg, Erlangen, Germany

[3]Department of Materials Science and Engineering, Friedrich-Alexander-Universität Erlangen-Nürnberg, Erlangen, Germany

[4]Department of Physics, Friedrich-Alexander-Universität Erlangen-Nürnberg, Erlangen, Germany

[5]Department of Materials Design and Innovation, University at Buffalo, NY, USA

## TOC

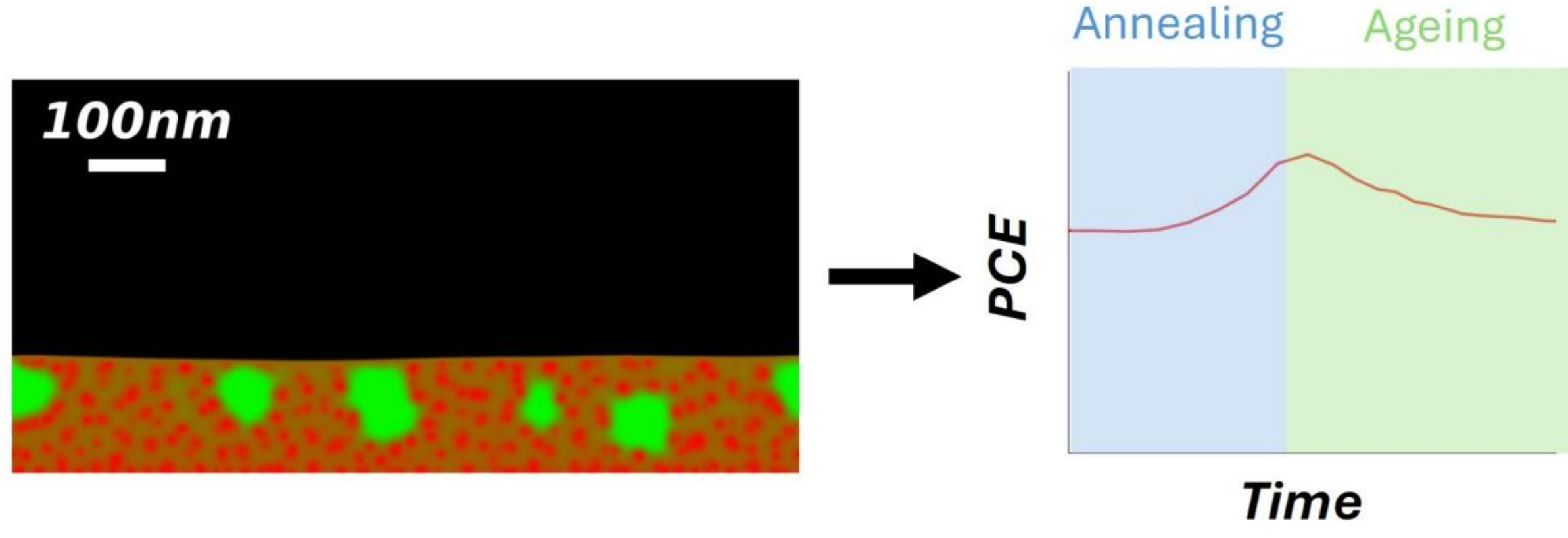

# Contents

# Abstract

Understanding the relationship between morphology and performance in organic solar cells is essential for developing devices that are both high-performing and resilient to aging. This work introduces a unique method capable of calculating the current-voltage (JV) curve of complex heterojunction morphologies containing up to five phases (donor amorphous, donor crystalline, acceptor amorphous, acceptor crystalline, mixed amorphous) with a very low computation time using morphology-aware descriptors of light absorption, exciton dissociation, non-geminate recombination and free charge carrier mobilities. The method is validated against Monte Carlo and 3D drift-diffusion simulations and applied to P3HT:PCBM and PM6:Y6 systems, shedding light on the physical compromises encountered to optimize device performance and lifetime. Finally, we show that the morphology-performance relationship is dependent on the materials system studied.

# 1. Introduction

The morphology of organic solar cell (OSC) absorber layers can significantly impact device performance. The energy conversion from light into electricity in OSC depends on the so-called "bulk-heterojunction" (BHJ) morphology. OSC absorber layers are typically composed of two semi-crystalline or crystalline materials (electron donor and electron acceptor) with multiple phases of very different properties and various morphologies found in BHJs. A situation with three to five phases is common, particularly for modern, fast-crystallizing donor polymers and non-fullerene acceptors. Yet, for sophisticated BHJ composites with three materials, the situation can be even more complex. The complexity of material phases, their distributions and the resulting variability in material properties and ultimately device performance significantly hinder progress. [1]

Understanding the relationship between BHJ morphology and device performance has been the focus of numerous studies aimed at optimizing the power conversion efficiency (PCE) of OSCs. Many studies focused on the diffusion length of excitons in organic materials (typically 10 nm to 50 nm). They highlighted the importance of small domain sizes (or equivalently, a high donor/acceptor interface density). By tailoring domain sizes, excitons can reach an interface within the BHJ, where they dissociate before geminate recombination. [2] [3] [4] [5] [6] [7] [8] [9] This is one example among many morphology-related design rules reported in the literature, several of which remain contradictory or system-dependent. It has also been shown that a large proportion of donor/acceptor mixed phase, where donor and acceptor molecules are close to each other, is beneficial for exciton dissociation. [7] But conversely, large domains (or equivalently a low donor/acceptor interface density) of high purity, or a low proportion of donor/acceptor mixed phase, help limiting non-geminate recombination of free charge carriers.[3,6,10,11] Mobilities are also expected to be at the highest for pure, crystalline donor and acceptor domains. [12,13] Last but not least, the BHJ must provide direct paths to the respective collecting electrodes, [12,14,15] either through the pure phases (acceptor for electrons, donor for holes) or through a mixed phase if the percolation threshold for charge transport is reached. [16–23]

Most of the knowledge above has been obtained by correlating selected BHJ characteristics to measurements of exciton dissociation, charge mobilities and/or recombination to the features of the device JV-curve. [24–26] The characterization of morphologies is challenging due to nanometer-scale features, low contrast between components and inherent limitations of microscopy. Hence, the correlations often remain partial. Recent advances in machine learning tools and optimization algorithms allowed for the analysis of large experimental databases to identify the morphology parameters that most strongly influence the device power conversion efficiency. Yet, they are also limited to easily accessible morphological descriptors, such as the film blend ratio and the crystallinities of donor and acceptor components. [27] [28] [29] [30] As a result, an overall precise and comprehensive picture of the morphology-properties relationship is still lacking.

Physics-based computational techniques offer the advantage of handling well-defined geometries and relating them to electronic properties. For a given morphology, exciton dissociation efficiency, non-geminate recombination, free charge carrier mobilities, or even the full JV curve, can be calculated. Multiple suitable morphological descriptors can be calculated from well-defined morphology and relate to the simulated properties. Multiple models, including 2D and 3D drift-diffusion, Monte Carlo, and master equation, have been used to calculate J-V curves of OSCs with sophisticated BHJ morphologies featuring two distinct amorphous phases or one amorphous and one crystalline phase. [2] [3] [4] [5] [6] [7] [8] [9] [10] [11] [12] [14] [15] [31] [32] [33] [34] [35] [36] [37] [38] [39] [40] [41] [42] [43] [44] [45] For co-continuous amorphous donor acceptor BHJ, the impact of domain sizes and purity on the exciton dissociation efficiency[9] [3] [5] [6] [8] [7] [37] and of domain sizes, purity and tortuosity on the charge carrier mobility[12] [14] [15] [42] has been confirmed. The impact of the domain size on non-geminate recombination has also been investigated.[6] [11] Nevertheless, these simulation methods are computationally very demanding and require advanced computational science proficiencies, hence, in practice, their use has not become widespread in the optoelectronics community. To tackle this problem, Wodo and coworkers developed a graph-based model to derive simple and quickly calculable descriptors of complex morphologies to relate them to electronic properties, providing a qualitative insight into cell performance.[1,43] They successfully identified descriptors for light absorption, exciton dissociation, and charge recombination, opening an avenue for morphology optimization. Yet, their work focuses on two-phase morphologies and is based on correlations between morphological features and performance metrics; hence, it still relies on computationally intensive device prediction.

To the best of our knowledge, previous simulation studies were limited to two-phase systems, mostly fully amorphous BHJ. One reason for this limitation is the lack of available realistic morphologies with more than two phases. A notable exception is the work of Li and coworkers, who generated (amorphous) three-phase BHJ morphologies from experimental characterizations and analyzed charge transport and recombination using the master equation approach.[44] Beyond this, our group has recently developed a phase-field framework that enables the simulation of BHJ formation in amorphous-crystalline donor-acceptor mixtures, allowing for the analysis of complex morphologies with two to five phases. [46] [47]

The present work introduces a simple, quick and easily accessible theoretical framework that consistently accounts for light absorption, charge dissociation, recombination and transport. It provides robust, physics-based relationships between morphology and properties. In relation to previously reported modeling approaches, it applies to morphologies with up to 5 phases (donor amorphous, donor crystalline, acceptor amorphous, acceptor crystalline, mixed amorphous). Morphology-aware descriptors are calculated from the morphology raw data, using a graph-based approach, for (1) exciton dissociation into collectible free charge carriers, (2) non-geminate recombination and (3) both free charge carriers' mobilities. Together with morphology-dependent optical properties obtained from spectral modeling, these descriptors are used as inputs to standard device physics modeling, employing a transfer-matrix formalism (TMF) coupled to a 1D drift-diffusion (DD) model. At the end, the BHJ morphology-dependent JV curve is calculated.

We apply our approach to morphologies obtained from physics-based phase-field simulations and to analytical morphologies featuring two to five crystalline and/or amorphous phases. Overall, this allows us to physically link the morphology of complex BHJ to the charge generation, transport, recombination rates, and the device performance. The computational time for a single simulation is below 1 minute, provided the morphology is available. We illustrate the power of the approach by analyzing the structure-property relationship for a set of morphologies, corresponding to various processing conditions and material systems. Finally, we show how the evolution of the BHJ morphologies under thermal loading can impact the device performance, paving the way for understanding the role of film post-treatment (thermal annealing) and intrinsic stability during cell operation on OSC efficiency.

The paper is organized as follows: The methodology is detailed in Section 1 and validated in Section 2. Section 3 presents the results obtained for a set of BHJ morphologies under different processing conditions, for both systems P3HT:PCBM and PM6:Y6, starting with the 'as cast' state and during subsequent thermal aging.

# 2. Methodology

Figure 1 shows the general workflow of the proposed morphology-performance analysis, from the morphology data of the active layer to the JV-curve of the solar cell. The primary input is the raw morphology data stored as a matrix of nodes. In this work, the input comes mostly from phase-field simulations of BHJ morphology formation. However, this is not restrictive, and data from other models or experimental measurements can be handled. First, the raw morphology data undergo a preprocessing step to assign phase and composition to each node. The composition is defined in terms of the donor and acceptor volume fractions, $\phi_d$ and $\phi_a = 1 - \phi_d$, respectively. The type of phase is defined as 'donor crystalline', 'acceptor crystalline', 'donor amorphous (nearly pure)', 'acceptor amorphous (nearly pure)' or 'mixed amorphous' (see Figure 3 and Figure 4). This data is used for calculating (i) the optical properties of the active layer with the help of a spectral model (see Sec. 2.1) and (ii) descriptors using a graph-based analysis (see Sec. 2.2). The descriptors capture characteristics of the morphology related to (1) the dissociation of excitons into (collectible) free charge carriers, (2) non-geminate recombination and (3) electron and hole mobilities. These morphology-aware descriptors, alongside the calculated optical properties, are fed into a 1D drift-diffusion device physics model to generate JV curves from which the fill factor ($FF$), the open circuit voltage ($V_{oc}$), the short circuit current ($J_{sc}$) and the power conversion efficiency ($PCE$) are extracted. The three main building blocks of this workflow are detailed below.

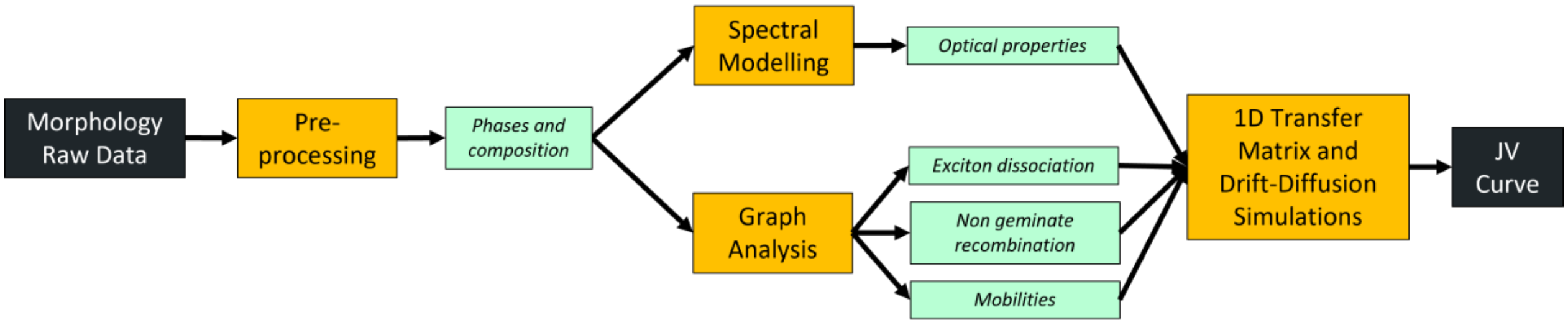

*Figure 1: Schematic representation of the workflow to calculate JV curves from a 2D morphology input. The figure highlights the flow of information between tools and models used (orange boxes), the calculated properties of the morphology (green boxes) and the input-output variables (black boxes).*

## 2.1. Morphology-dependent optical properties using spectral modeling

Light absorption is the first step in photovoltaic conversion in organic solar cells. It is characterized by the absorption spectrum $S(\lambda)$ of the active layer. The absorption spectrum of an organic molecule arises from the electronic transitions from the ground state to allowed excited states. Spano has proposed a spectral model for the optical properties of organic semiconducting molecules that accounts for excitonic intermolecular coupling, exciton-phonon coupling, and disorder.[48] [49] Building on this model, Lüer and coworkers developed a machine learning surrogate model, trained with high-throughput experimental data, that links an experimentally acquired material system absorption spectrum to morphological features (blend ratio and crystallinities), using Gaussian deconvolution.[50] [51] Applied to OSC BHJ, this leads to the identification of four spectra standing for the donor in the crystalline/aggregated state, $S_{d,cr}(\lambda)$, the donor in an amorphous state, $S_{d,am}(\lambda)$, the acceptor in crystalline/aggregated state $S_{a,cr}(\lambda)$, and the acceptor in an amorphous state $S_{a,am}(\lambda)$. In this paper, we use the surrogate model to determine the spectrum of the morphology given the volume fractions and crystallinity of the constituting phases. The spectrum is computed as a linear combination of four spectra,

$$S(\lambda) = \overline{\phi_d}\,\overline{\chi_d}S_{d,cr}(\lambda) + \overline{\phi_d}(1-\overline{\chi_d})S_{d,am}(\lambda) + \overline{\phi_a}\,\overline{\chi_a}S_{a,cr}(\lambda) + \overline{\phi_a}(1-\overline{\chi_a})S_{a,am}(\lambda), \tag{1}$$

where $\overline{\phi_d}$ and $\overline{\phi_a}$ are the average donor and acceptor volume fractions, associated with the donor/acceptor blend ratio in the photoactive layer. $\overline{\chi_d}$ and $\overline{\chi_a}$ are the crystallinity of the donor and acceptor averaged over the whole morphology:

$$\overline{\chi_d} = \sum_{d,cr}\phi_d \Big/ \sum_{BHJ}\phi_d \tag{2}$$

for the donor (and a similar expression for the acceptor), whereby index $d, cr$ denotes the set of nodes of the crystalline donor phase and index $BHJ$ denotes the set of nodes on which the morphology is defined. Figure 2 illustrates the spectral signatures of the amorphous and crystalline phases of PM6 and Y6 used in this work for the calculation of the morphology spectrum.

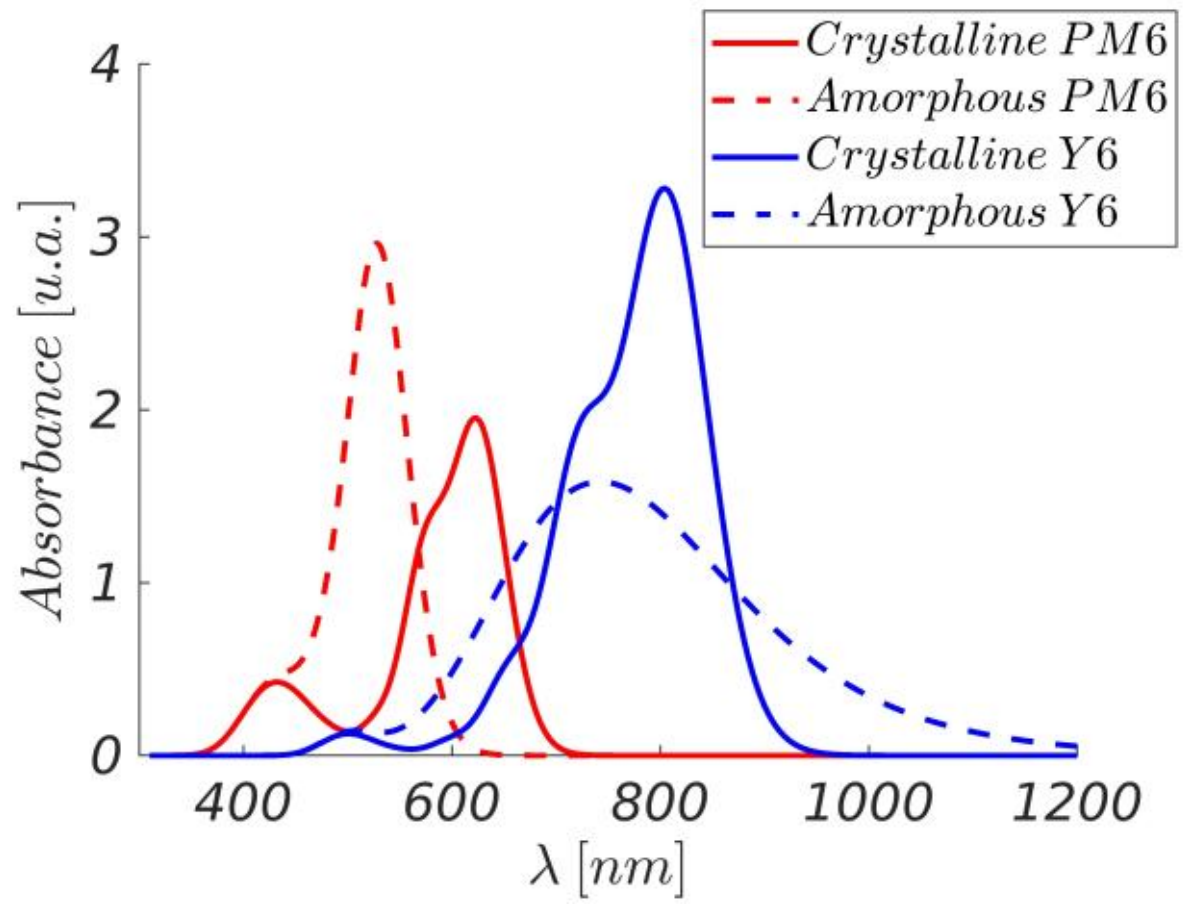

*Figure 2: Absorption spectrum of crystalline PM6, amorphous PM6, crystalline Y6 and amorphous Y6 obtained from the approach described in [51].*

In order to calculate light absorption in the solar cell, the TMF model [52] requires the complex refractive index $n(\lambda) + ik(\lambda)$ of the active layer. It is determined from the absorption spectrum of the BHJ $S(\lambda)$ in the following way: the imaginary part $k(\lambda)$ of the refractive index is calculated using [52]

$$k(\lambda) = \frac{\lambda S(\lambda) l\,n(10)}{4\pi h} \tag{3}$$

where $h$ is the thickness of the active layers from which the spectra $S(\lambda)$ are obtained. The real part of the refractive index is calculated using the Kramers-Kronig equation[53–56]:

$$n(\lambda) = 1 + \frac{2}{\pi}\wp\int_0^{\infty} \frac{k(\lambda')}{\lambda'\left(1-\left(\frac{\lambda'}{\lambda}\right)^2\right)} d\lambda' = 1 + \frac{2}{\pi}\wp\int_{\lambda_{min}}^{\lambda_{max}} \frac{k(\lambda')}{\lambda'\left(1-\left(\frac{\lambda'}{\lambda}\right)^2\right)} d\lambda' + C \quad (4)$$

Here, $\wp$ corresponds to the Cauchy principal value of the integral. The constant $C$ compensates the inevitable mismatch between the experimental refractive index and the values obtained by the Kramer-Kronig equation, which is due to the truncated integration of the absorption coefficient $k(\lambda')$ over the experimentally measured spectral range. It is obtained by adjusting the level of the calculated $n(\lambda)$ to available ellipsometry data.

## 2.2. Morphology-aware electronic properties using graph-based analysis

Graph-based descriptors capture morphology-dependent electronic properties from the time the exciton is formed until the charges are collected at the electrodes. Below, we provide an overview of the processes involved in the current generation, their relationship with morphology and the methodology used to capture their dependence on morphology. Once photons are absorbed, an exciton is formed. To collect the corresponding energy, the exciton needs to be dissociated into free charge carriers, which typically occurs at the interface between the donor and acceptor materials. If the exciton does not reach that location, geminate recombination takes place. In our simplified five-phase morphology, the locations where an exciton can dissociate are (a) the interfaces between the crystalline or amorphous acceptor and donor phases and (b) the mixed amorphous phase (acceptor and donor molecules close enough to provide an opportunity for dissociation). Once dissociated, both electrons and holes require a path to their respective electrodes. Otherwise, the corresponding energy gets lost, e.g., an electron is trapped in an isolated island of acceptor material. Additionally, charge carriers may recombine in regions where they can meet each other, where both donor and acceptor materials are available (interfaces and mixed phase). Calculating descriptors for exciton dissociation, charge transport and recombination thus requires, as a preliminary step, the careful definition and identification of functional ***regions in the BHJ morphology, namely:***

- The charge-transport phases are defined as the regions in the morphology where free charge carriers can travel. The *electron transport phase (ETP)* consists of the acceptor crystalline phase, the nearly pure acceptor amorphous phase, and the mixed phase. Similarly, the *hole-transport phase (HTP)* comprises the donor crystalline phase, the nearly pure donor amorphous phase, and the mixed amorphous phase. This means that charge transport of both electrons and holes is assumed to take place in the mixed phase. Thereby, the mixed amorphous phase and the nearly pure donor/acceptor amorphous phases are identified using percolation thresholds for electron and hole transport, respectively, [16–23] in the following way: the percolation thresholds are defined as the critical acceptor (resp. donor) volume fraction ${\phi_a}^{crit}$ (resp. ${\phi_d}^{crit}$) below which electron (resp. hole) transport is not possible anymore. Building on this, the donor phases (forbidden for electron transport) are regions with acceptor volume fractions $\phi_a < {\phi_a}^{crit}$, the acceptor phases (forbidden for hole transport) are regions with donor volume fractions $\phi_d < {\phi_d}^{crit}$, and the mixed phase corresponds to regions where ${\phi_d}^{crit} \le \phi_d \le 1 - {\phi_a}^{crit}$.

- Building on the ETP and the HTP, we define the effective transport phases in the following way (see Figure 3 and Figure 4): first, the *effective electron transport phase* (*EETP)* is restricted to the part of the ETP that is connected to the cathode, and the *effective hole transport phase* (*EHTP)*, to the part of the HTP that is connected to the anode. Therefore, only charge carriers in the effective transport phases can contribute to the current. We further constrain the effective hole (electron) transport regions of morphology by identifying these domains that are directly adjacent to the effective electron (hole) transport phase. In this way, we identify the regions of morphology where *both* charges have pathways towards their respective electrodes and from which excitons can diffuse to the dissociation location. This is a critical step to ensure that the analysis of the whole photovoltaic process (charge generation, transport, recombination) is restricted to those contributions only that eventually lead to current generation.
- Finally, we define the region *common to both effective transport phases (CETP,* see Figure 3 and Figure 4). Following the logic above, only the excitons dissociated in this region will lead to a collectible current, because only from there can both dissociated charge carriers be extracted from the BHJ. The CETP obviously contains the portion of the mixed phase that belongs to both EETP and EHTP. In addition, the interfaces between the (crystalline or amorphous) acceptor phase belonging to the EETP and the (crystalline or amorphous) donor phase belonging to the EHTP are assumed to be regions from where both charge carriers have available pathways to the electrodes. Since the interface itself is 'off lattice', we define the nodes of the EETP and of the EHTP neighboring these interfaces as part of the CETP, Provided the domain sizes are large as compared to the grid spacing, this simplification has a negligible impact on the results. In practice, most of the interfaces are diffuse and feature composition gradients. As a result, a thin mixed phase is present, most of the time, between pure donor and acceptor regions.

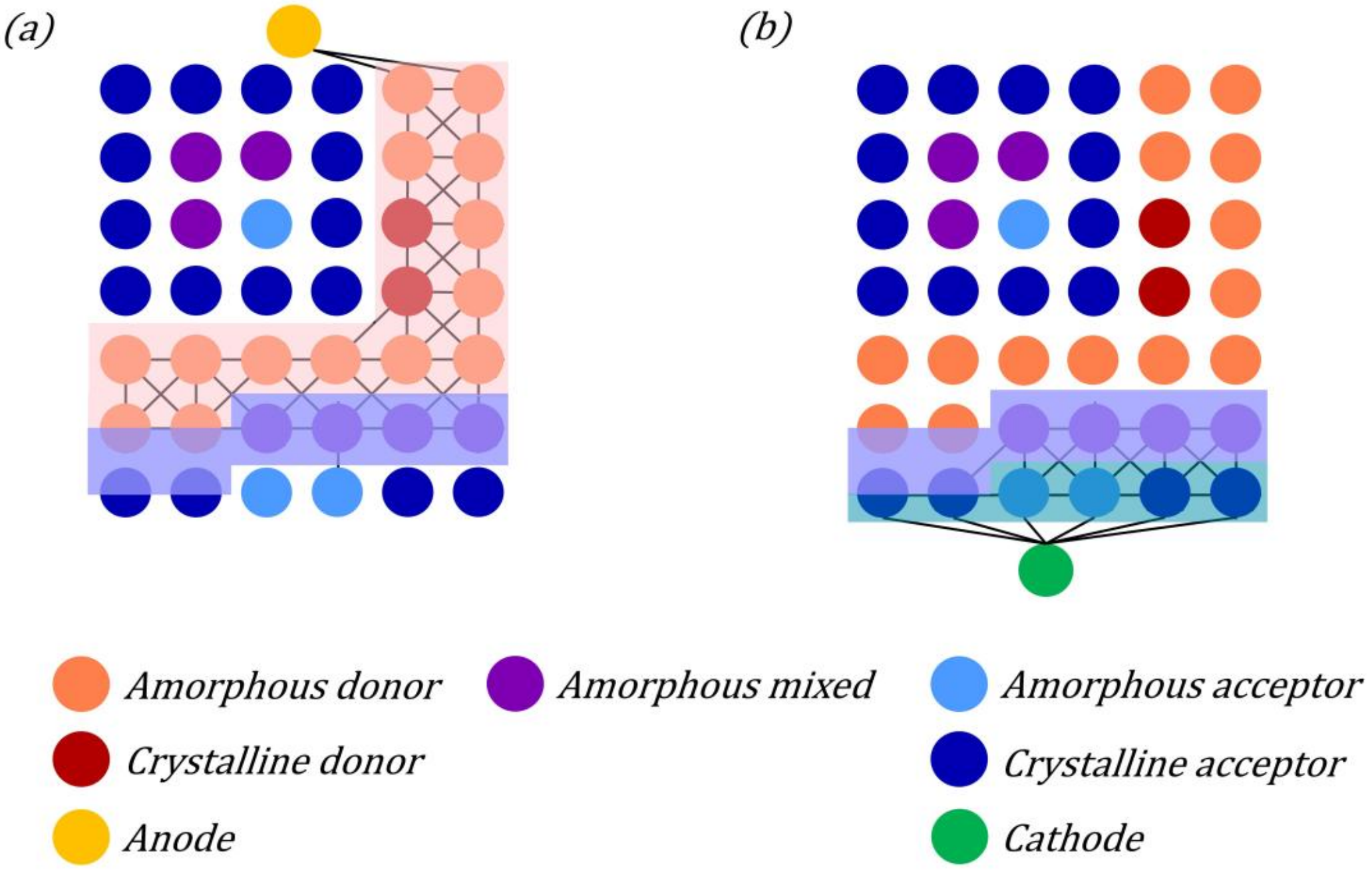

*Figure 3: Use case of the graph-based model and connected component labeling on a morphology. The graph-based representation with the connected path of (a) the effective hole*

*transporting phase to the anode and (b) the effective electron transporting phase to the cathode are illustrated. The crystalline and amorphous phases of the donor and acceptor nodes are represented in dark red and light red, and dark blue and light blue, respectively. The amorphous mixed nodes, the anode and the cathode are represented in purple, yellow, and green, respectively. The black segments correspond to the different connections between the nodes of the EETP or the nodes of the EHTP. The salmon, turquoise, and purple background colors correspond to the EHTP nodes not in the CETP, the EETP nodes not in the CETP, and the CETP nodes, respectively.*

The identification of the EETP, EHTP and CETP requires a connectivity analysis between grid points. These functional regions are determined using a graph-based approach, in which each voxel is represented as a vertex labeled with the phase type to which it belongs, and neighboring vertices are connected by edges weighted by the distance between them.[1] [14] [57] [58] A connected components labeling algorithm is used to identify the nodes of the EETP, EHTP and CETP. Figure 3 illustrates the graph-based representation of a given toy model morphology, along with its corresponding EETP, EHTP, and CETP.

The gradual refinement of regions with assigned phases into effective transport regions, and ultimately into common effective transport regions, enables forecasting whether all steps of photovoltaic processes can be completed. In this paper, we use them to calculate the following descriptors:

### 2.2.1. Descriptor for exciton dissociation

We first define a descriptor for excitons dissociation as the average "efficiency of exciton dissociation into collectible free charge carriers" $\overline{\eta_d}$, which we call "exciton dissociation efficiency" for simplicity. It takes into account not only the effectiveness of morphology for excitons dissociation, resulting from their diffusion to regions where it can dissociate, but also the ability of both generated electrons and holes to be collected at the electrodes. The dissociation efficiency for an exciton generated at a given location $\eta_d$ is determined in the following way: (a) $\eta_d = 0$ if it is generated at a node belonging to neither EETP nor EHTP, because such locations actually correspond to islands transporting a given type of free charge carrier fully surrounded by the transport phase of the other type of free charge carriers. Therefore, the free charge carriers generated in the islands upon exciton dissociation have no pathways for being extracted. (b) $\eta_d = 1$ for an exciton generated in the CETP, because donor and acceptor molecules are available for dissociation, and because a pathway to the electrodes is available for both charge carriers, by definition of the CETP. (c) Excitons generated at nodes that are in the EETP or the EHTP, but not in the CETP, can actually be dissociated into collectible free charge carriers if they manage to reach the CETP. Thus, for each such node, we calculate the distance to the closest CETP node, leveraging the graph-based description of the BHJ. Then, we assume that the dissociation probability decreases exponentially with this distance, with the typical length scale being the exciton diffusion length. The local exciton dissociation efficiency reads as follows:

$$\eta_d = \begin{cases} 1 & in\ CETP \\ e^{\frac{-d_d}{L_d}} & in\ EHTP\ but\ not\ CETP \\ e^{\frac{-d_a}{L_a}} & in\ EETP\ but\ not\ CETP \\ 0 & out\ of\ EETP\ and\ EHTP \end{cases} \tag{5}$$

Thereby, $L_d$ and $L_a$ are the donor and acceptor exciton diffusion lengths, respectively, and $d_d$ and $d_a$ are the closest distances to the CETP for an exciton generated by light absorption in the donor and in the acceptor, respectively. (e) Following the exciton diffusion pathways to the CETP, excitons may enter a mixed phase region not belonging to the CETP. They are assumed to dissociate here with 100% probability, but charge carriers are not both collectible, so that they are lost for current generation. To take this into account, we assume the distances $d_d$ and $d_a$ to be infinite in such situations.

Figure 4 illustrates the field of local exciton dissociation efficiency for a five-phase morphology obtained from phase-field simulations. Finally, the exciton dissociation efficiency for the whole BHJ $\overline{\eta_d}$ is the average of the local dissociation efficiencies over the $N$ nodes of the BHJ:

$$\overline{\eta_d} = \frac{1}{N} \sum_{BHJ} \eta_d \tag{6}$$

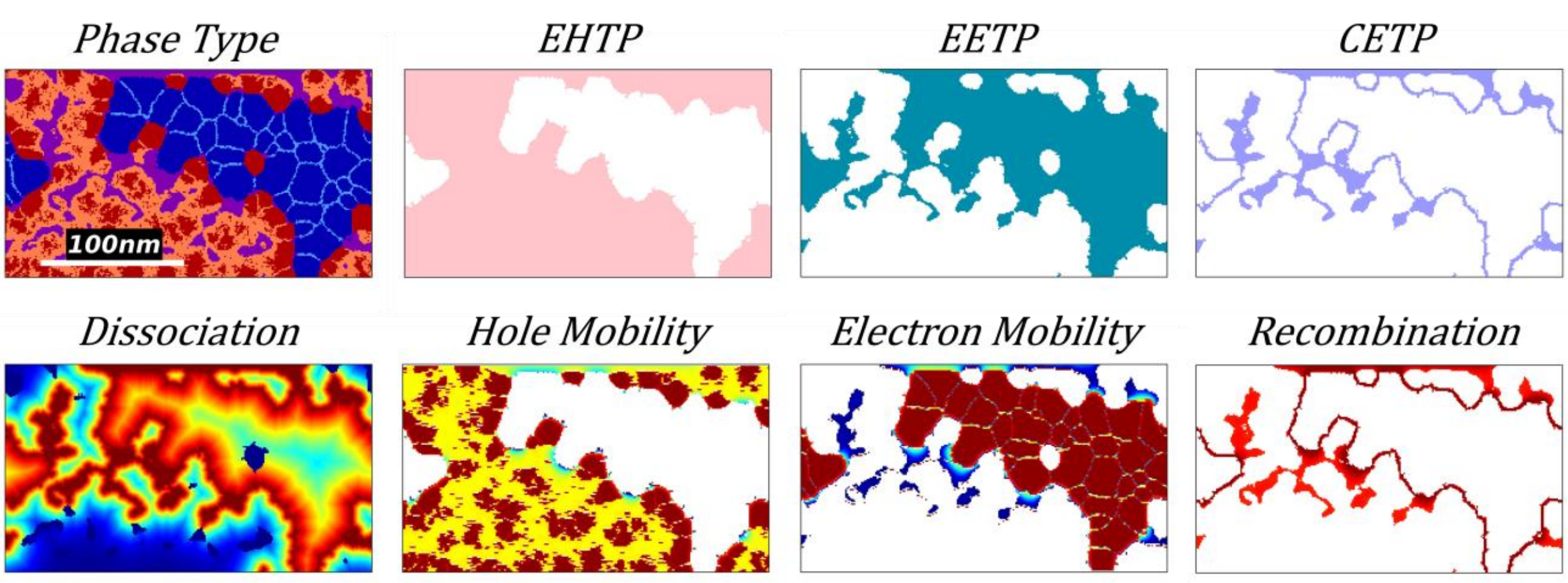

*Figure 4: Field values of the morphology-aware descriptors for a five-phases morphology. (Top) Functional regions, from left to right: phase type (donor amorphous in light red, donor crystalline in dark red, acceptor amorphous in light blue, acceptor crystalline in dark blue, mixed amorphous in purple), effective hole transport phase region (EHTP, salmon), effective electron transport phase region (EETP, turquoise), region common to both effective transport phase (CETP, violet). (Bottom) Fields of morphology-aware electronic properties, from left to right: exciton dissociation efficiency into collectible free charge carriers $\eta_d$, hole mobility $\mu_h$, electron mobility $\mu_e$ and recombination $k_r$. The values scale linearly from 0 (dark blue) to 1(dark red) for the dissociation efficiency, and logarithmically from $10^{-3}$ (dark blue) to 1 (dark red) for mobilities and recombination. The corresponding average values for the whole morphology are $\overline{\eta_d} = 0.60$, $\overline{\mu_h} = 2.93 \cdot 10^{-2}$, $\overline{\mu_e} = 1.94 \cdot 10^{-2}$, $\overline{k_r} = 8.45 \cdot 10^{-2}$.*

### 2.2.2. Descriptor for non-geminate bimolecular recombination

The main idea here is that only the recombination of collectible free charge carriers is relevant to device behavior, and that these carriers can meet only in the CETP, by definition. Therefore, we define a local recombination prefactor $k_r$ only in both EETP and EHTP, which is non-zero only in the CETP. At the nodes of the CETP, following statistical physics considerations similar to those underlying the Flory-Huggins theory of mixing,[59] $k_r$ is assumed to depend on the local composition according to $k_r = 4\phi_d\phi_a$. Microscopically, $\phi_d\phi_a$ represents the probability that a donor molecule (where holes can be found) is neighboring an acceptor molecule (where electrons can be found) and therefore it stands for the likelihood of electron-hole encounter at the considered node. The factor 4 ensure a normalization to $k_r = 1$ for $\phi_d = \phi_a = 0.5$. Figure 4 exemplifies the field of local recombination descriptor for the same morphology as before. Finally, the recombination descriptor for the whole BHJ $\overline{k_r}$ is the average of the local recombination descriptors over the effective transport phases:

$$\overline{k_r} = \frac{1}{N_{EETP \cup EHTP}} \sum_{EETP \cup EHTP} k_r = \frac{1}{N_{EETP \cup EHTP}} \sum_{CETP} 4\phi_d\phi_a \tag{7}$$

Thereby, $N_{EETP \cup EHTP}$ is the number of nodes in either the EETP or the EHTP.

### 2.2.3. Descriptors for Shockley-Read-Hall recombination (SHR)

Three descriptors are defined in a manner very similar to the previous descriptor by filtering nodes that meet a given criterion, and by weighting their contributions based on the local volume fraction. To capture the effect of morphology on SHR recombination, three types of traps are considered: traps near the LUMO, traps near the HOMO and mid-gap traps. It is assumed that only electrons can be captured by traps close to the LUMO. Hence, we filter nodes belonging to the EETP, and the probability of recombination at these nodes is proportional to the donor volume fraction. This leads to a SHR recombination descriptor for the morphology-dependent number of LUMO-edge traps $\overline{N_{te}}$ equal to the average impurity of the EETP:

$$\overline{N_{te}} = \frac{1}{N_{EETP}} \sum_{EETP} \phi_d \tag{8}$$

Similarly, hole trapping through HOMO-edge traps can occur in the EHTP regions, leading to the descriptor

$$\overline{N_{th}} = \frac{1}{N_{EHTP}} \sum_{EHTP} \phi_a\,. \tag{9}$$

Mid-gap traps can be present in either EETP or EHTP, and can capture either electrons traveling in the EETP or holes traveling in the EHTP, so that we define a SHR mid-gap trap recombination descriptor $\overline{N_{teh}}$ as follows:

$$\overline{N_{teh}} = \frac{1}{N_{EETP} + N_{EHTP}} \left( \sum_{EETP} \phi_d + \sum_{EHTP} \phi_a \right) \tag{10}$$

### 2.2.4. Descriptors for charge carrier mobility

Only the mobility of collectible charge carriers is relevant for device physics modeling, so that the analysis is restricted to the EETP for electrons and to the EHTP for holes, respectively. Let us describe the approach for electron mobility; the calculation for holes is essentially the same. In order to define the local mobility at a given node $\mu_e$ we assume the charge transport to be promoted along the electric field gradient, namely the vertical direction, towards the cathode. This means that charge carriers may hop to the next layer of nodes towards the cathode, and we assume they may hop either to the nearest neighboring node (exact vertical hopping) or to the second-nearest neighboring nodes. We assume that the hopping probability to the various neighbors depends on (a) the type of phase in the target node, (b) the composition of the target nodes, and (c) the angle with the vertical direction for a hop to the target node. On top of that, it is assumed that (d) the transition from a crystalline phase to an amorphous phase is hindered by an energetic barrier and that (e) hopping out of the EETP is not allowed. [13,60] Ultimately, we assume that local mobility is related to the most likely hopping opportunities to neighboring nodes. More specifically we have

$$\mu_e = max_{Neighb,EETP}\left[{\phi_a}^2\left(\chi_a \mu_{a,cr} + (1 - \chi_a)\mu_{a,am}\right)\left(\delta_{cram} p_e + (1 - \delta_{cram})\right)|co\,s(\alpha)|\right], \tag{11}$$

where, the maximum value of mobility is selected from the neighborhood constituted of target nodes, with the acceptor volume fraction $\phi_a$, crystallinity $\chi_a$ ($\chi_a = 0$ for a node in an amorphous phase and $\chi_a = 1$ in a crystalline phase), and local mobility $\mu_{a,cr}$ (electron mobility in the pure crystalline acceptor) and $\mu_{a,am}$ (electron mobility in the pure amorphous acceptor ($\mu_{a,am} < \mu_{a,cr}$)). Finally, $\delta_{cram}$ is an indicator of a crystalline-to-amorphous hopping ($\delta_{cram} = 1$ for such a jump, $\delta_{cram} = 0$ otherwise), $p_e$ is a penalty factor for a jump from the crystalline to the amorphous phase ($p_e \leq 1$), and $\alpha$ is the angle between the hopping direction and the vertical direction. In Eq. (11), the function $max_{Neighb,EETP}$ denotes that the maximum value over the nearest and next nearest neighbors of the EETP is taken. Intuitively, the local mobility is the highest between crystalline, pure node and crystalline, pure node exactly in the vertical direction. By analogy, the mobility descriptor for holes is defined as

$$\mu_h = max_{Neighb,EHTP}\left[{\phi_d}^2\left(\chi_d \mu_{d,cr} + (1 - \chi_d)\mu_{d,am}\right)\left(\delta_{cram} p_h + (1 - \delta_{cram})\right)|co\,s(\alpha)|\right]. \tag{12}$$

Figure 4 exemplifies the field of local hole and electron mobility descriptors for the same morphology as before.

Finally, the average velocity along the drift trajectories is the harmonic average of the local values. This is because drift-diffusion equations actually reduce to advection-like equations, with a velocity $\mu\vec{E}$, if the generation, recombination and diffusion terms are neglected. Also, we assume that the electric field is homogeneous in the active layer, which is a reasonable first-order approximation. For simplicity, we also approximate the charge-carrier trajectories to be vertical during the effective transport phases. Under these conditions, we use the harmonic average of the local mobilities to determine the average mobility value along vertical streamlines during the effective transport phase. Then, we take the arithmetic mean of all the streamline-averaged mobilities over the lateral dimensions to obtain the average mobilities $\overline{\mu_e}$ and $\overline{\mu_h}$ for the whole BHJ:

$$\overline{\mu_e} = \frac{1}{N_{st,EETP}} \sum_{j=1}^{N_{st,EETP}} \left( \frac{N_{st,EETP,j}}{\sum_{i=1}^{N_{st,EETP,j}} \frac{1}{\mu_e}} \right) \tag{13}$$

Thereby, $N_{st,EETP}$ is the number of vertical streamlines in the EETP, and $N_{st,EETP,j}$ is the number of nodes in the streamline $j$. Similarly for holes, we have

$$\overline{\mu_h} = \frac{1}{N_{st,EHTP}} \sum_{j=1}^{N_{st,EHTP}} \left( \frac{N_{st,EHTP,j}}{\sum_{i=1}^{N_{st,EHTP,j}} \frac{1}{\mu_h}} \right). \tag{14}$$

Note that the mobility strongly depends on the transport direction in organic molecular crystals encountered in common donor and acceptor materials, the mobility being much larger in the $\pi - \pi$ stacking direction. This may be an important effect that has not been considered because the phase-field morphologies analyzed in the present work do not provide any reliable information about crystal orientation. Nevertheless, this limitation will be released in the near future, and the implementation of a mobility descriptor depending on the transport direction in the crystals is straightforward within the proposed framework.

To conclude, the table below summarizes the symbols defined for the morphology analysis discussed above, along with the values chosen in the current work.

| Parameter | Notation | Value, PM6-Y6 |
|---|---|---|
| Absorption spectrum, donor amorphous | $A_{d,am}$ | See Figure 2 |
| Absorption spectrum, donor crystalline | $A_{d,cr}$ | See Figure 2 |
| Absorption spectrum, acceptor amorphous | $A_{a,am}$ | See Figure 2 |
| Absorption spectrum, acceptor crystalline | $A_{a,cr}$ | See Figure 2 |
| Percolation threshold for hole transport | ${\phi_d}^{crit}$ | *0.1* |
| Percolation threshold for electron transport | ${\phi_a}^{crit}$ | *0.1* |
| Exciton diffusion length in donor | $L_d$ | *10 nm* |
| Exciton diffusion length in acceptor | $L_a$ | *35 nm* |
| Hole mobility in the pure crystalline donor | $\mu_{d,cr}$ | *1* |
| Hole mobility in the pure amorphous donor | $\mu_{d,am}$ | *0.1* |
| Electron mobility in the pure crystalline acceptor | $\mu_{a,cr}$ | *1* |
| Electron mobility in the pure amorphous acceptor | $\mu_{a,am}$ | *0.1* |
| Penalty factor for an electron hop from crystalline to amorphous phase | $p_e$ | *0.01* |
| Penalty factor for a hole hop from crystalline to amorphous phase | $p_h$ | *0.01* |

*Table 1: Symbols used for morphology analysis of PM6-Y6 and of P3HT-PCBM bulk-heterojunctions.*

## 2.3. Morphology-aware 1D drift diffusion device physics model

To simulate the JV curve associated with a given active layer morphology, two models are used: (i) a classical one-dimensional device physics model based on a TMF optical module and then (ii) the drift-diffusion equations.[52] [61] [62] [63] As stated before, the complex refractive index is directly used as an input of the TMF model, which allows for calculating the local exciton generation rate $G_{ex}$. The dissociation, mobility and recombination descriptors, $\overline{\eta_d}$, $\overline{\mu_h}$, $\overline{\mu_e}$, $\overline{k_r}$, $\overline{N_{te}}$, $\overline{N_{th}}$ and $\overline{N_{teh}}$ capture the characteristics of the BHJ nanoscale morphology in relation to the device physics. To account for intrinsic properties of materials (energetics, molecular arrangement, defect density…), we define the following input parameters:

$$\eta_d{}^{DD} = \overline{\eta_d}\eta_d{}^{i} \tag{15}$$

$$\mu_h{}^{DD} = \overline{\mu_h}\mu_h{}^{i} \tag{16}$$

$$\mu_e{}^{DD} = \overline{\mu_e}\mu_e{}^{i} \tag{17}$$

$$k_r{}^{DD} = \overline{k_r}k_r{}^{wc} \tag{18}$$

$$N_{te}{}^{DD} = \overline{N_{te}}N_{te}{}^{wc} \tag{19}$$

$$N_{th}{}^{DD} = \overline{N_{th}}N_{th}{}^{wc} \tag{20}$$

$$N_{teh}{}^{DD} = \overline{N_{teh}}N_{teh}{}^{wc} \tag{21}$$

Here, $\eta_d{}^{i}$, $\mu_h{}^{i}$, $\mu_e{}^{i}$ are the exciton dissociation efficiency in an ideal BHJ, the hole mobility in the crystalline donor, and the electron mobility in the crystalline acceptor. $k_r{}^{wc}$ is the prefactor for bimolecular recombination in the 'worst case' situation, e.g., a fully homogeneous, amorphous mixed morphology for a 1:1 donor acceptor blend. Similarly, $N_{te}{}^{wc}$, $N_{th}{}^{wc}$, $N_{teh}{}^{wc}$ are the prefactor for SHR recombination in the 'worst case' situation, with fully impure transport phases. Note that the above parameters are defined as the product of the morphology-aware descriptor and the corresponding parameter of the non-BHJ-morphology-related contribution.

The parameters $\eta_d{}^{DD}$, $\mu_h{}^{DD}$, $\mu_e{}^{DD}$, $k_r{}^{DD}$, $N_{te}{}^{DD}$, $N_{th}{}^{DD}$ and $N_{teh}{}^{DD}$ are input to the drift-diffusion model in the following way: $\eta_d{}^{DD}$ is used as a multiplier to the local, inhomogeneous generation rate at each mesh point; $k_r{}^{DD}$ can be either a prefactor for Langevin recombination, e.g. the recombination rate is $R = k_r{}^{DD}(q\,(\mu_e{}^{DD} + \mu_h{}^{DD})/\varepsilon)n_e n_h$, whereby $n_e$ and $n_h$ are the electron and hole densities, $\varepsilon$ the dielectric permittivity of the material, and $q$ the elementary charge, or a non-Langevin bimolecular recombination factor, e.g. $R = k_r{}^{DD}n_e n_h$.

# 3. Validation of morphology-aware electronic descriptors

To validate our simplified approach for evaluating morphology-dependent electronic properties, the values for exciton dissociation, bimolecular recombination, and mobilities have been calculated using Equations *6*, *7*, *13* and *14* above and compared with previously published simulation results. We identified four studies basing on advanced physics-based simulation methods (2D or 3D drift-diffusion, Monte-Carlo, Master equation for well-defined morphologies[2] [3] [4] [5] [6] [7] [9] [11] [12] [14] [15] [37] [42] where the simulation results can be compared to the present approach. [4] [6] [7] [13] They discuss different morphological descriptors and types of morphology, but all these studies treat only two-phase morphologies (either amorphous demixed donor-acceptor BHJ or amorphous-crystalline single-material morphologies). Yang and coworkers calculated the exciton dissociation efficiency for planar, chessboard, planar-mixed, and homogeneous heterojunctions of different thicknesses using Monte Carlo simulations.[4] Using 2D and 3D drift-diffusion simulations, Kodali investigated the exciton dissociation efficiency and the non-geminate recombination in two-phase amorphous co-continuous morphologies resulting from spinodal decomposition, with variable domain sizes.[6] Groves performed Monte Carlo simulations of exciton dissociation in a similar morphology, depending on domain size and purity.[7] [8] The dependence of the mobility on crystallinity and crystal size in a single semi-crystalline material has been reported by Geng, based on Master Equation simulations.[13] We reproduced the morphologies investigated by these authors and calculated the morphology descriptors using the present approach. All results are in line with the published simulation data (see Supporting Information, Section 1.1). To extend this valuable but limited validation database to a broader range of morphologies, we generated analytical, simple donor-acceptor morphologies with up to five phases, featuring amorphous and/or crystalline regions. We calculated and verified the morphology-dependent exciton dissociation, hole and electron mobilities and bimolecular recombination (see Supporting Information, Section 1.2).

# 4. Results

## 4.1. Morphology-dependent JV curves and their evolution upon thermal loading

To illustrate the interest and the power of the proposed approach, we apply the method to a set of various organic photoactive layer BHJ morphologies, for which we identify the phase distribution, calculate the optical and electronic properties, and finally the JV curve of the corresponding solar cells, as described above. We gain insight from the simulation to establish the morphology-performance relationship and explain how and why these morphologies result in different JV curves.

The morphologies are taken from a database of 2D phase-field simulations of BHJ formation upon solution processing available in our group. The morphology set corresponds to as-cast morphologies resulting from various processing conditions and/or thermodynamics and kinetic properties of the drying mixture. The reader is referred to previous work for more details on the process-structure relationship.[46] [64] The chosen morphologies feature three to five phases and differ in various ways (see the phase pattern of the morphologies in Figure 5 and the average composition, crystallinity and the phase amount in the Supporting Information, Section 3.1). Morphology M1 consists of four phases with small, partly crystalline donor domains in an amorphous mixed phase matrix, featuring an isolated region of amorphous donor. Morphology M2 also consists of four phases but with large donor crystals surrounded by a mixed phase, and an acceptor amorphous phase which contains few acceptor crystals. Morphology M3 consists of five phases, with the donor being less crystalline than in M2, and the acceptor being more crystalline. Morphology M4 consists of only three phases with bilayer-like morphology, with an amorphous acceptor layer below an amorphous donor layer containing donor crystals. Morphology M5 is similar to M2 and M3, but with a different balance between the amount of the donor and acceptor crystalline phases and a different spatial organization. Morphology M6 is a three-phase morphology with well-balanced crystallinity of donor and acceptor, whereby small elliptical crystals are dispersed in an amorphous matrix.

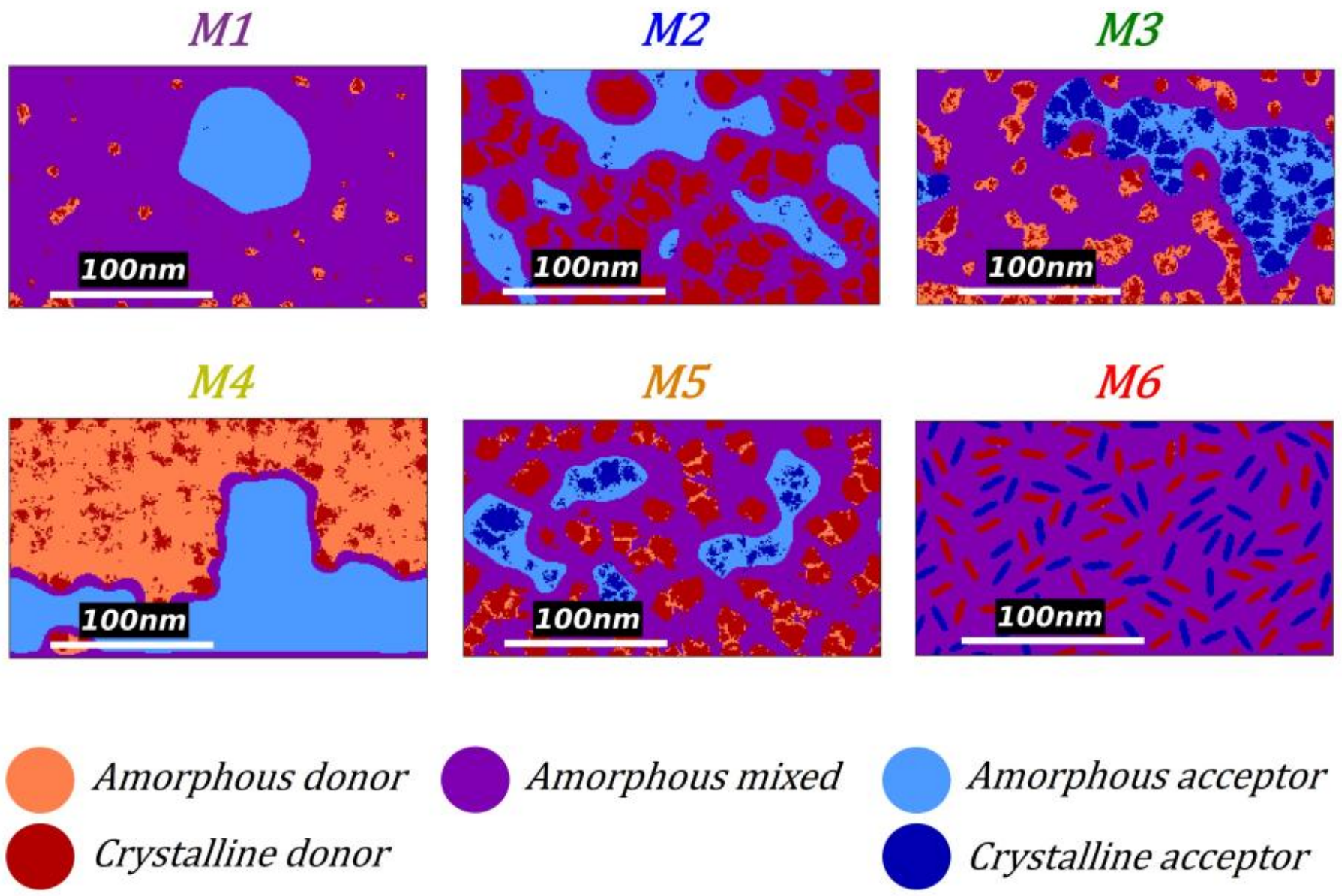

*Figure 5: Visualization of the phase-type and their distribution in the as-cast film morphologies selected for calculation of morphology-dependent JV curves. The morphologies are obtained from phase field simulations of BHJ formation during processing.*[46] [64]

It is assumed that these morphologies can be considered as representative of as-cast, sub-optimal PM6:Y6 photoactive layers of solar cells for which the JV curves have to be simulated. It is not the topic of this work to discuss the validity of this assumption. For now, it allows to calculate the absorption spectrum of PM6:Y6 blends organized according to these morphologies, as well as exciton dissociation, mobility and recombination descriptors, using model parameters representative of the PM6:Y6 blend (see Table 1 and the supporting Information, Section 2). In order to calculate JV curves representative of PM6:Y6 solar cells, the morphology-independent properties ${\eta_d}^i$, ${\mu_h}^i$, ${\mu_e}^i$, ${k_r}^{wc}$ and ${N_{teh}}^{wc}$ of the PM6:Y6 material system are obtained in the following way: the JV curve data obtained experimentally for a high-performance solar cell is fitted by drift-diffusion modeling. Thereby, without loss of generality, the bimolecular recombination is assumed to be non-Langevin for simplicity, and only the SHR trap recombination associated with mid-gap traps is considered. The values for the mobilities, bimolecular and SHR recombination rate and dissociation efficiency are obtained from the fit, and it is assumed that the values ${\mu_h}^i$ and ${\mu_e}^i$ are about one decade above the fitted electron and hole mobilities, and the ${k_r}^{wc}$ and ${N_{teh}}^{wc}$ one decade below, in line with the idea that the fitted solar cell is an excellent but not ideal one. Finally, ${\eta_d}^i$ is equal to 1. More details on this calibration method can be found in Supporting Information, Section 2. Note that SHR trap recombination is found to be negligible in the considered example and will not be discussed further in the remainder of the paper.

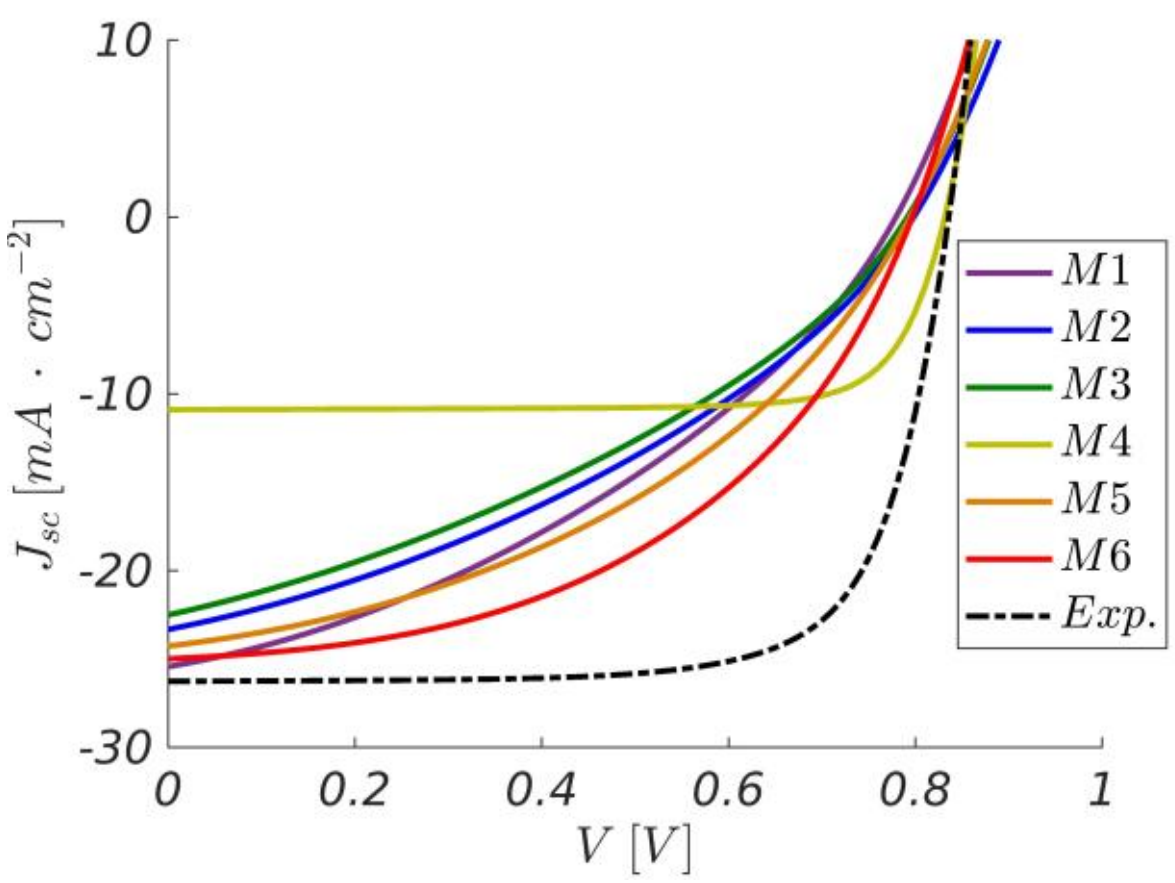

*Figure 6: Simulated JV curves of organic solar cells under 1 sun illumination with a photoactive layer made of PM6:Y6, for the morphologies shown in Figure 5 (colored curves), together with the experimentally obtained JV curve (see Ref. [65], black curve).*

The JV curves for devices with the selected morphologies are shown in Figure 6, together with the experimental JV curve. The $FF$, $V_{oc}$, $J_{sc}$ and $PCE$ values can be found in the Supporting Information, Section 3.1. We observe very significant variations of the JV curve with the BHJ morphology. As expected, since the virtual morphologies correspond to as-cast layers and are not optimized, the best device is the real one, by far. Note that 'ideal' morphologies such as thin pillar structures provide better performances as compared to the real cell (See Supporting Information, Section 1.2). The search for morphologies that provide best performances while being experimentally accessible will be the topic of future investigations.

These six morphologies were used to simulate their evolution under thermal loading using our phase-field framework, to qualitatively represent what happens to the BHJ during a thermal annealing step or during aging. Improving the lifetime of OSC remains a major challenge, as the impact of thermal loading on intrinsic BHJ stability is not yet fully understood. Overall, all selected morphologies experience further "phase separation" upon thermal loading. More specifically, demixing of the amorphous phase as well as donor and acceptor crystallization proceed further, so that the amount of mixed phase decreases, and the amount of crystalline donor and acceptor phases increases. On top of that, crystal growth and coarsening of the demixed amorphous phases lead to a size increase of all domains in the BHJ, both amorphous and crystalline. More details about this time evolution can be found in previous work.[46] Note that such an evolution, and consequently the results below, cannot be generalized for all donor-acceptor blends, since it is strongly dependent on the thermodynamic and kinetic properties of the mixtures.[66] [67]Nevertheless, the present approach allows us to calculate time-dependent JV-curves for all selected morphologies, and Figure 7 shows the associated short circuit current, fill factor, open circuit voltage and power conversion efficiency. While $J_{sc}$ decreases with time for all morphologies, the fill factor increases. The evolution of $V_{OC}$ is very limited and will thus not be discussed further in the following. As a result of the balance between $J_{sc}$ and the fill factor, the efficiency may decrease, increase, or have a non-monotonic behavior depending on the initial morphology. If the thermal loading is thought of as a thermal annealing step, this illustrates that, depending on the as-cast morphology, thermal annealing can, of course, be beneficial for "morphology optimization". However, as is commonly observed in practice, the annealing time must be properly adjusted for each as-cast morphology to avoid excessive morphological evolution. In the following, the reasons for these variations are analyzed in detail.

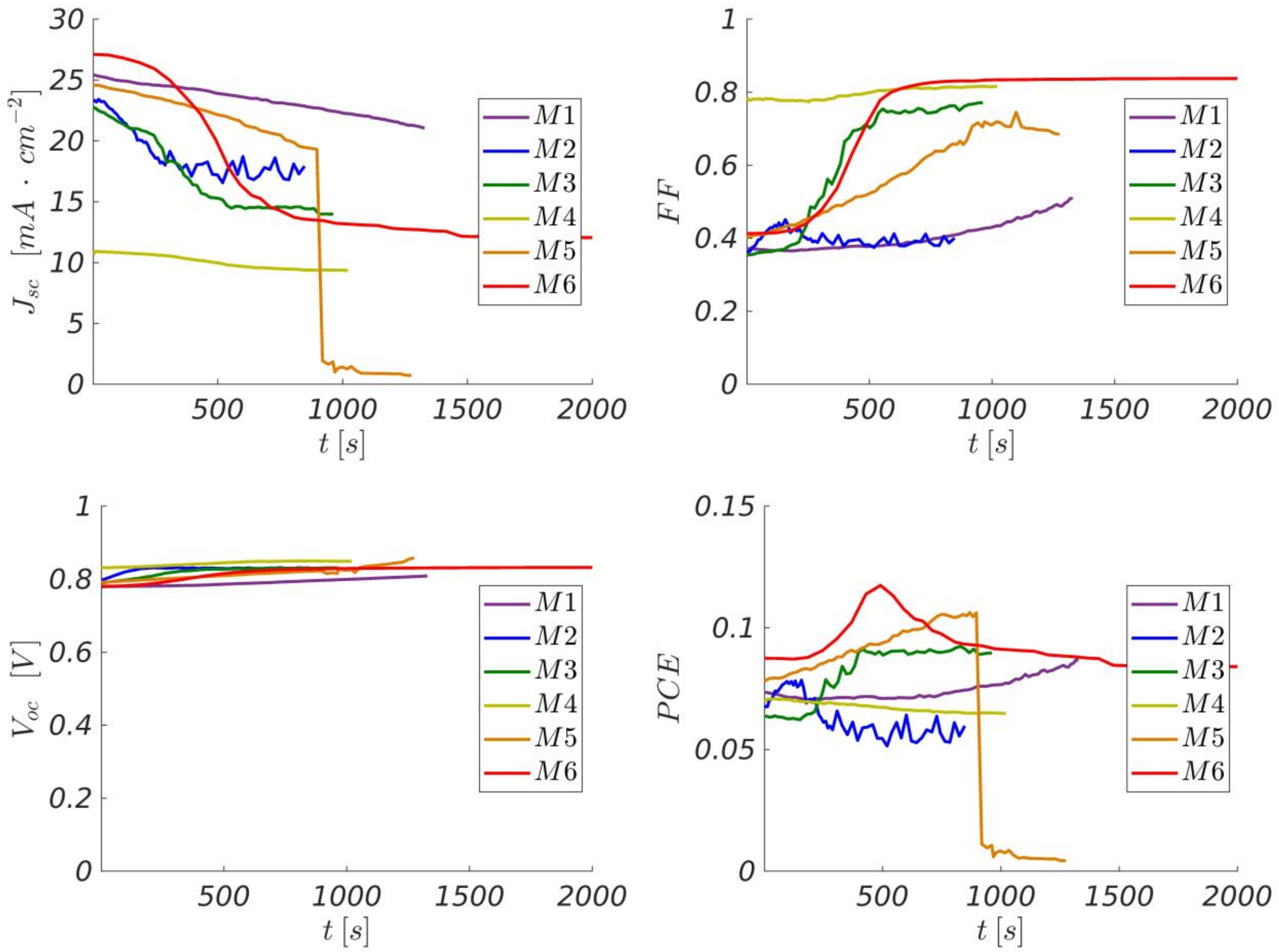

*Figure 7: Short circuit current ($J_{sc}$), fill factor ($FF$), open circuit voltage ($V_{OC}$) and power conversion efficiency ($PCE$) evolution upon thermal annealing followed by aging for a PM6:Y6 solar cell.*

As expected for reasonably good solar cells, the $J_{sc}$ variations are fully related to the generation rate $G$ of variations of the number of (collectible) free charge carriers. The proportionality $J_{sc} \sim G$ is nicely recovered (see Figure 8, left). This results from changes in the exciton dissociation efficiency ${\eta_d}^{DD}$, together with changes in the exciton generation rate $G_{ex}$, namely $G \sim {\eta_d}^{DD} \cdot G_{ex}$. In the present case, the variations of exciton generation rate are much smaller than the variations of exciton dissociation efficiency, as can be seen by the large horizontal shift between the curves tracking $G$ and those tracking $G_{ex}$.

Neher and co-workers have suggested that the fill factor results from a balance between charge collection and recombination.[68] This led them to propose a relationship between the fill factor and a single descriptor

$$\alpha = \frac{qh^2}{2kT}\sqrt{\frac{{k_r}^{DD} G}{{\mu_e}^{DD} {\mu_h}^{DD}}} \tag{22}$$

using a modified, empirical Green equation

$$\begin{cases} FF = \frac{u_{OC} - ln(0.79 + 0.66 u_{OC}{}^{1.2})}{1 + u_{OC}} \\ u_{OC} = \frac{qV_{OC}}{kT(1+\alpha)} \end{cases}. \tag{23}$$

Equation *22* is expected to hold for the case of balanced mobilities. For imbalanced mobilities, various weighting laws of both mobilities have been proposed.[69] [70] Interestingly, in the present study, we found that using the smallest of both mobilities provides the best match with the modified Green equation, and we thus use the slightly modified descriptor $\alpha'$ for the fill factor:

$$\alpha' = \frac{qh^2}{2kT}\sqrt{\frac{k_r{}^{DD}G}{min(\mu_e{}^{DD}, \mu_h{}^{DD})^2}} \tag{24}$$

Figure 8 (right) shows the dependency of the fill factor on $\alpha'$ for all considered morphologies, at any time of ageing, which nicely follows the modified Green equation. Note that the horizontal shift between the simulation data and the Green equation is due to our drift-diffusion modeling choices, which account for much greater complexity (3 layers, series resistance...) than the simulations of Neher and co-workers. Neglecting these effects and performing simulations as in Neher et al., we find a perfect overlap of simulation data points and the Green equation (see Supporting Information Section 3.2).

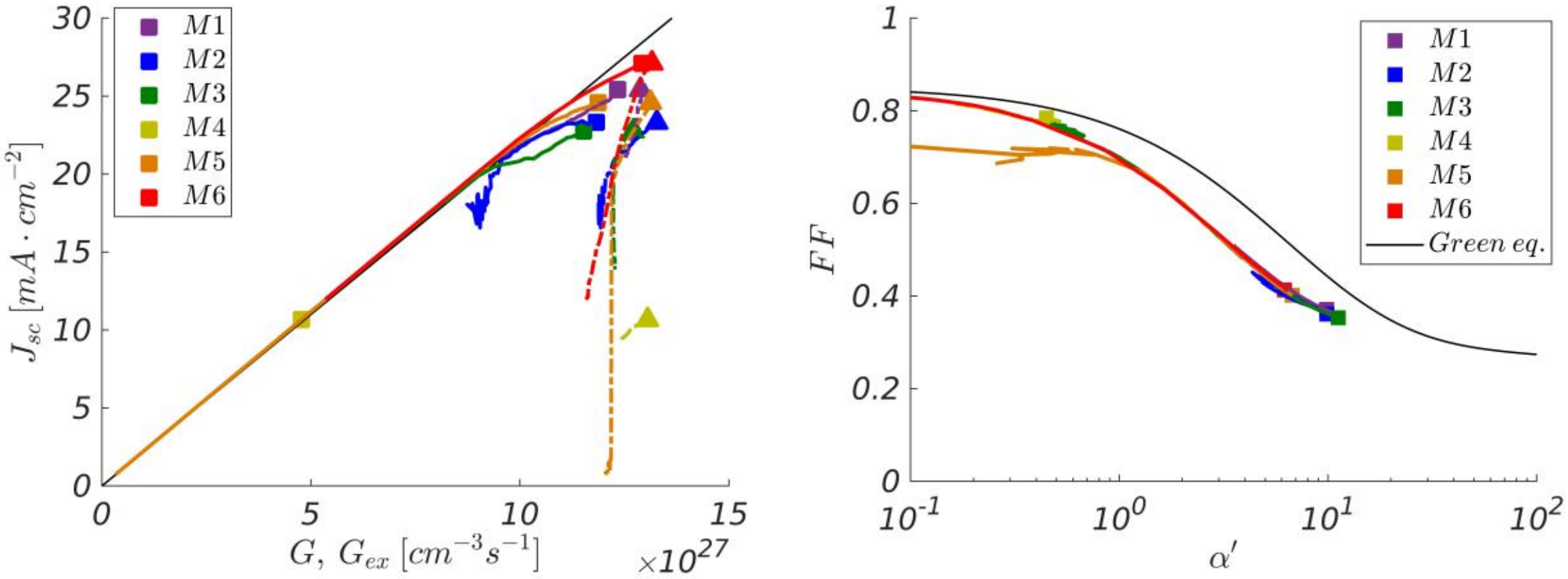

*Figure 8: Relationship between optoelectronic performance and charge generation, recombination and collection efficiency. Each curve represents the time evolution for a given morphology, whereby the large symbols mark the as-cast morphology. (Left) $J_{sc}$ plotted against the exciton generation rate $G_{ex}$ (triangles) and the free charge carrier generation rate $G$ (squares). (Right) FF plotted against $\alpha'$, Equation 24. The horizontal shift between the simulation points and the Green equation is due to contributions to the fill factor that are unrelated to the absorber layer and are not accounted for by Equation 24 (see Supporting Information Section 3.2).*

Based on this pre-analysis and using Equations (*15*-*18*), the BHJ morphology can be fully related to solar cell performance, taking advantage of $J_{sc} \sim G \sim \overline{\eta_d} \cdot G_{ex}(S(\lambda))$ and of $FF = f(\alpha') = f\left(\frac{\overline{k_r}G}{min(\overline{\mu_e}, \overline{\mu_h})^2}\right)$. Therefore, in the following, we focus on how the morphology-aware properties $S(\lambda)$, $\overline{\eta_d}$, $\overline{k_r}$, $\overline{\mu_e}$ and $\overline{\mu_h}$ defined in Section 2 depend on the morphology.

## 4.2. Analysis of the morphological origin of the optoelectronic properties

The evolution of the exciton generation rate $G_{ex}$ with annealing time for each morphology is shown in Figure 9 (left). Overall, $G_{ex}$ decreases with time, and its variation is limited to less than 15%. Since the blend ratio is fixed for a given morphology (60% vol. donor for M1 to M5 and 50% vol. donor for M6), the time evolution is obviously related to the evolution of the absorption spectrum with donor and acceptor crystallinity (see Equation *1*). As stated before, both donor and acceptor crystallize with time. This results in a redshift of the absorption spectrum and a strong reduction in Y6 absorption in the infrared for $\lambda > 900nm$. Additionally, this redshift is responsible for the disappearance of the PM6 amorphous peak at 550 nm and the appearance of a crystalline peak at 650 nm, shifting the PM6 absorption to a region of the AM1.5 light spectrum where the intensity is slightly lower. Overall, the PM6:Y6 absorption spectrum becomes unfavorable at high crystallinity, compared to what would be optimal for proper absorption of the AM1.5 spectrum (Supporting Information Section 3.3). Figure 9 (left) shows that exciton generation is optimal for a nearly amorphous acceptor and a roughly 50% crystalline donor. Regarding as-cast morphologies, this explains why the exciton generation is higher for M2 and M5. Note that light absorption is greater in M6 due to the more favorable blend ratio. Upon annealing, the time evolution trajectories of the considered morphologies unfortunately bypass this optimal crystallinity ratio.

In the current example, the exciton dissociation efficiencies are much more sensitive to the BHJ morphology than the exciton generation rate, with dissociation efficiencies ranging from nearly 0 to nearly 1 (see Figure 9, right). Exciton dissociation efficiencies are globally larger with a larger amount of amorphous mixed phase $\phi_{AmM}$ because excitons systematically dissociate in the mixed phase. But this is not the end of the story: the BHJ spatial organization and the domain sizes play a crucial role. The as-cast morphologies can be classified into three categories: M4 results in a very low exciton dissociation efficiency, simply because it is a bilayer-like structure with layer thicknesses much larger than the exciton diffusion lengths, which promotes considerable exciton deactivation. Morphologies M1 and M6 feature dissociation efficiencies in collectible free charge carriers $\overline{\eta_d}$ close to 1, because (a) they feature an amorphous mixed matrix where generated excitons can be dissociated and which ensures that all free charge carriers can travel towards the electrodes, and (b) the pure donor and acceptor crystalline inclusions are small enough for most of the excitons generated there to reach the mixed phase where they separate. Note that for M1, despite the large central Y6 droplet, exciton recombination is limited thanks to the large exciton diffusion length of Y6. Morphologies M2, M3 and M5 also feature an amorphous mixed phase, which provides pathways to the electrodes from everywhere in the morphology (e.g., all nodes belong either to the EETP or to the EHTP, there are no isolated islands). However, they also feature pure donor and acceptor domains that are large as compared to the exciton diffusion lengths, so that a significant amount of the excitons recombines because they cannot reach the CETP. Note that due to the smaller diffusion length in PM6, for M2 and M5, most of the excitons are lost in the pure PM6 domains, even though they are smaller than the pure Y6 domains. For M3, most of the excitons are lost in the large Y6 central region.

Regarding the time evolution under thermal loading, the dissociation efficiency decreases over time for all morphologies, as domain sizes increase and the amount of mixed phase decreases. This time, evolution is very limited for M4, because the bilayer topology remains nearly unchanged. For M1 and M2, the evolution is quite limited as well, because the morphology evolves in such a way that donor and acceptor domains remain small enough and/or remain at least connected to the CETP by short pathways. On the contrary, M3 and M6 develop large donor and acceptor domains, some of which become isolated islands, which explains the more pronounced decrease in exciton dissociation efficiency. Finally, the dissociation efficiency for the morphology M5 features an abrupt drop after some time. This can be understood as follows: upon thermal loading, a strong phase separation between pure donor and acceptor domains arises, so that the CETP (the region providing pathways for both charge carrier types to their respective electrodes) is eventually made of thin veins between these domains. Unfortunately, a donor phase develops at the bottom of the morphology (cathode side), so that a large part of the CETP is connected to the cathode by only a couple of veins. The catastrophic drop in dissociation efficiency corresponds to the time when this donor layer nearly covers the entire cathode, cutting the last connecting veins and isolating a large portion of the BHJ from the cathode (see Supporting Information Section 3.4). This illustrates how minor morphological changes can lead to drastic changes in optoelectronic properties. In particular, slight morphological changes may significantly affect the connecting pathways to the electrodes, especially when they consist solely of interfaces, thereby altering the amount and geometry of effective transport phases and isolated domains. $J_{sc}$, $FF$, and $PCE$ fluctuations, as visible in Figure 7, are also related to reconfigurations of connecting pathways upon minimal BHJ evolution.

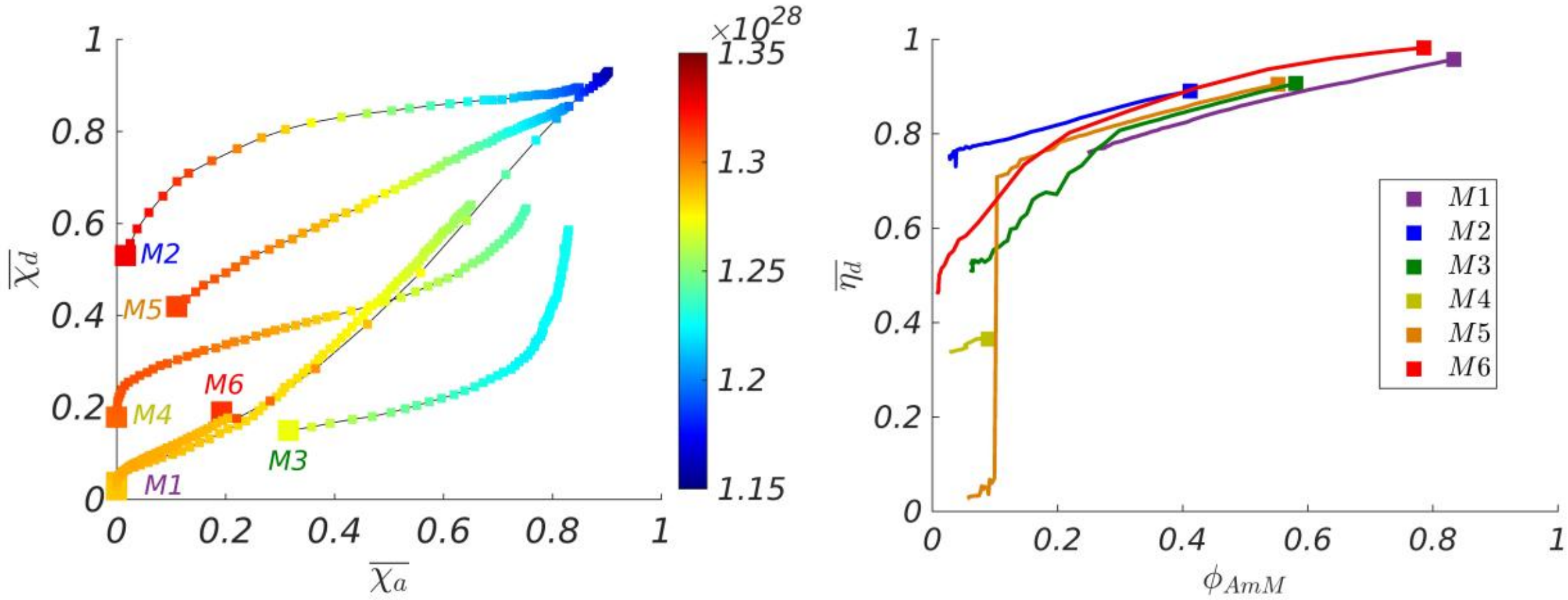

*Figure 9: Impact of the BHJ morphology on the generation rate. Each curve represents the time evolution for a given morphology, whereby the large symbols mark the as-cast morphology. (Left) evolution of the exciton generation rate (color code) with average acceptor crystallinity $\overline{\chi_a}$ and donor crystallinity $\overline{\chi_d}$. (Right) Average exciton dissociation efficiency $\overline{\eta_d}$ depending on the volume fraction of amorphous mixed phase $\phi_{AmM}$ in the BHJ.*

Next, we focus on recombination properties. As shown in Figure 10 (left), the average bimolecular recombination descriptor $\overline{k_r}$ is approximately proportional to the proportion of amorphous mixed phase $\phi_{AmM}$, because in the current examples, the amorphous mixed phase is the main contributor to the CETP, which is the region where electrons and holes may recombine. Deviations from this approximation are larger at small amounts of amorphous mixed phase, because the contribution of interfaces to the CETP is more significant. Obviously, from Equation 7, the correlation between the recombination descriptor and the ratio of the CETP volume fraction to the volume fraction of effective transport phases is excellent (see Supporting Information Section 3.5). The composition of the CETP only plays a secondary role. This reflects and confirms the idea that charge transport should occur in separate, pure donor and acceptor phases, rather than in mixed phases, to limit non-geminate recombination. As a result, for the as-cast morphologies, the amount of recombination is higher for M1 and M6, where the "phase separation" is the most limited. Note that for the bilayer morphology M4, the recombination region is limited to the thin interface between both layers, as expected, providing excellent recombination properties. Upon thermal loading, as phase separation proceeds, the amount of the mixed phase decreases, and the donor and acceptor phases are purified, thereby decreasing the recombination probability across all morphologies.

To finish with, the impact of BHJ morphology on the smallest mobility $min(\overline{\mu_e},\overline{\mu_h})$ is shown in Figure 10 (right). As stated before, increasing the limiting mobility is key for improving the fill factor. In the analyzed morphologies, the electron mobility is the limiting one most of the time (full lines in Figure 10, right), except for morphologies M3 and M4 at the end of the thermal evolution (dotted lines in Figure 10, right). The value of each morphology descriptor is a subtle balance between purity, crystallinity and energy penalty for hopping from a crystalline to an amorphous phase, remember Equations *11* to *14*. Pure, highly crystalline phases are obviously expected to favor large mobility values, and this explains the global trend for mobility increase upon thermal loading. However, this holds only at sufficiently high crystallinities: if frequent hopping between amorphous and crystalline regions is required for vertical transport, mobility is penalized at low crystallinity (see Supporting Information Section 1.1). This is, for instance, why hole mobility is smaller than electron mobility in the bilayer morphology M4, even though the donor layer is partially crystalline and the acceptor layer is fully amorphous (see Figure 5). On top of this, the average mobility strongly depends on the topology of the effective transport phases. For instance, both morphologies, M2 and M6, exhibit very high crystallinities at the end of the thermal evolution. However, the electron mobilities differ significantly. In morphology M2, electrons have to travel through thin mixed-phase veins in between donor crystals to reach the cathode, resulting in a low mobility. In morphology M6, electrons can travel through large acceptor "highways" nicely separated from donor domains. All in all, comparing the mobility values for the six investigated morphologies requires a detailed step-by-step analysis that goes beyond the scope of this paper. However, we wish to emphasize that obtaining highly crystalline, highly pure donor and acceptor phases is not sufficient to achieve large, balanced mobilities. The optimization of the domains spatial organization is crucial for improving the mobilities.

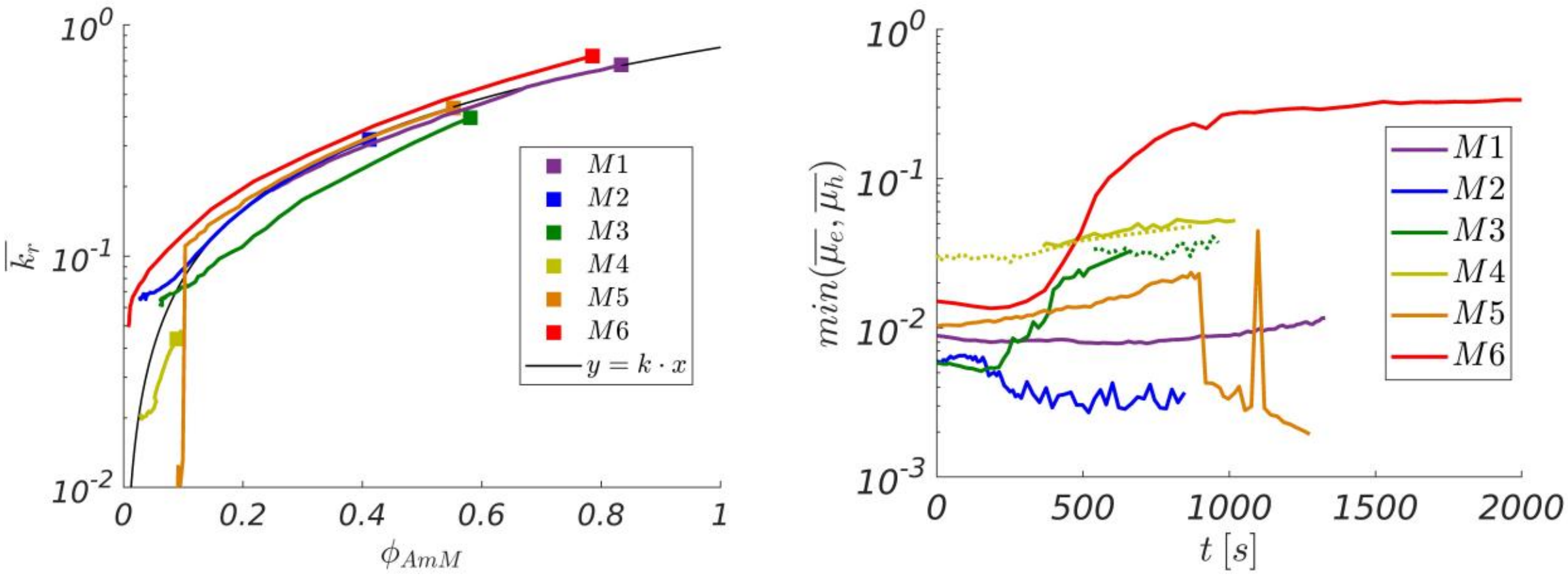

*Figure 10: Impact of the BHJ morphology on the recombination and transport properties. (Left) Average bimolecular recombination rate depending on the volume fraction of amorphous mixed*

*phase $\phi_{AmM}$ in the BHJ. Each curve represents the time evolution for a given morphology, whereby the large symbols mark the as-cast morphology. (Right) Smallest mobility depending on time for each morphology (full line: $\overline{\mu_e} < \overline{\mu_h}$ and dotted line: $\overline{\mu_e} \geq \overline{\mu_h}$).*

To summarize this analysis, photovoltaic performance and its evolution result from a balance between numerous effects. It has been shown that the commonly expected effects are recovered (the role of blend ratio and crystallinity on absorption properties, the role of domain sizes on exciton dissociation efficiencies, the positive impact of phase separation on recombination properties, the compromise between exciton dissociation efficiency and recombination, the role of purity and crystallinity on mobilities...). Beyond this, the present framework allows us to take into account the crucial role of the spatial organization of the BHJ (pathways for exciton diffusion to the CETP, pathways for charge carriers to the collecting electrodes...), and, finally, to balance all effects together. If morphology is simple, quick conclusions can be drawn for the structure-property relationship. For instance, for the bilayer-like morphology M4, the performance is determined by a high fill factor on the one hand, due to the clear separation of the electron and hole transport phases with a minimal interface layer in between, and a poor generation rate due to an exciton dissociation limited to a thin region around this interface. Moreover, the performance is quite stable under thermal loading, largely because the BHJ geometry is relatively stable. If the BHJ morphologies are more complex, summarizing the whole balance might be more challenging. For instance, the as-cast morphologies M2, M3 and M5 look quite similar, and so are their optoelectronic properties. Nevertheless, their slightly different donor and acceptor crystallinities result in noticeable differences in the as-cast JV curves (see Figure 6). Moreover, their evolutions upon aging strongly differ. For M3, crystallization, together with demixed-phase coarsening and purification, results in a drop in free charge-carrier generation (due to both reduced light absorption and lower exciton dissociation efficiency) and an increase in fill factor (due to reduced bimolecular recombination and improved mobilities). M5 initially experiences a similar evolution, but a catastrophic drop of the exciton dissociation efficiency due to the BHJ geometry evolution results in a huge performance breakdown. For M2, despite a qualitatively similar morphology evolution, the fill factor does not increase because the limiting mobility does not increase, which is due to electron transport being impeded in thin veins between donor domains. In contrast, the as-cast morphology of M6 differs significantly from that of M3. However, the fill factor is very similar throughout the whole thermal loading. The property difference comes from charge carrier generation: thanks to a more favorable blend ratio, and to an initially smaller crystallinity (both in terms of amount and domain sizes), $J_{sc}$ is initially higher for M6 than for M3, and the $J_{sc}$ drop takes place later. As a result, the performance of M6 is better and features a maximum upon thermal loading, which is not present for M3.

On top of these purely morphological considerations, it should be emphasized that the results regarding photovoltaic performance depend on the material's optoelectronic properties taken into account through the parameters of the morphology analysis gathered in Table 1 and the values of the morphology-independent properties ${\eta_d}^i$, ${\mu_h}^i$, ${\mu_e}^i$ and ${k_r}^{wc}$. Results using different material parameters for the morphology analysis, and also using materials and device parameters for a completely different solar cell based on a P3HT-PCBM blend, are summarized in the Supporting Information Section 4. At the end, this also confirms that optimal morphology is *a priori* not the same for different donor-acceptor material systems, and that the processing route to reach the best possible morphology should also be, in principle, adapted to each donor-acceptor blend. Overall, this work illustrates the whole complexity of the topic, and therefore, we do not pretend to draw definitive conclusions on the structure-performance relationships at this stage. It is rather a proof of concept which shows that the current approach can handle all aspects of the problem, enabling very fast morphology screening and extensive mechanistic analysis. Thus, it is a tool that might strongly contribute to the rationalization of structure-property relationships and to BHJ morphology optimization in the future.

# 5. Conclusions

With this work, we introduced a new method to study the morphology-performance relationship of OSC. The approach relies on a topological analysis of the BHJ morphology and on the basic physics of the optoelectronic processes involved in power conversion, namely light absorption, free charge carrier generation, transport and recombination. Compared to previously published approaches limited to two or at best three phases, it can handle complex BHJ morphologies containing up to five phases, which is crucial for standard, modern OPV material blends. The method was successfully validated against previously published Monte-Carlo, master equation and 3D drift-diffusion simulations, although future, more extensive and precise comparison with such advanced methods is highly desirable. In addition, it was used for a detailed comparative analysis of the evolution of photovoltaic performance across six different morphologies under thermal loading. The analysis results in a full, precise understanding of the structure-performance relationship.

Thanks to its low computational cost (5-50 seconds to obtain the JV curve of a 2D morphology on a standard laptop), it opens the way to quick, systematic screening of morphology parameters in the near future, enabling the establishment of robust structure-property relationships and rational optimization of the BHJ morphology. As demonstrated in the present work, it can be applied to process-aware phase-field simulations of BHJ morphology formation. This opens a new opportunity to deal with the full process-structure-property relationship from a physical perspective, and therefore to answer the question how to fabricate the best possible morphology, depending on the chosen donor-acceptor material system. Moreover, provided some data on the BHJ evolution during device operation is available (for instance, from PF simulations), it can also be used to understand and control the intrinsic stability of OSC under thermal loading, and eventually contribute to OSC lifetime improvement. On the contrary, as a fast structure-property connector, it can also be used as a new back-mapping tool to support the identification of BHJ morphology through optoelectronic property measurements.

Finally, due to its relative simplicity and the low computational demand, it is easily usable for anyone with basic optoelectronic education, without the need of deep computational science skills. Moreover, even though morphologies from phase-field simulations have been used in the present work, the proposed structure-property method can be applied to morphologies obtained from other sources, such as other theoretical approaches or even experimental measurements, provided spatial information on composition and crystallinity is available. As a result, it is well-suited for usage in different research groups, whatever the nature of the available data, and could contribute to the democratization of a mechanistic approach of the process-structure-property relationship in the OPV community, and beyond this, in the organic electronics community.

# 6. Conflict of interests

There are no conflicts to declare.

# 7. Data availability

The simulation data used for this article is publicly accessible on Zenodo (see DOI:10.5281/zenodo.20272551).

# 8. Acknowledgements

The authors acknowledge financial support by the German Research Foundation (DFG, project HA 4382/14-1, project BR 4031/21-2 and within the Collaborative Research Center 'ChemPrint', project 538767711, CRC 1719), the European Commission (H2020 Program, Project 101008701/ EMERGE), the Helmholtz Association (SolarTAP Innovation Platform), and the profile center "FAUsolar" of the Friedrich-Alexander-Universität Erlangen-Nürnberg.

# 1. Validation of morphology-aware electronic descriptors

## 1.1. Comparison to published Monte Carlo, master equation and 2D-3D drift diffusion simulation results.

### 1.1.1. Exciton dissociation for layered morphologies of binary blends

Yang and coworkers performed Monte-Carlo simulations to calculate the exciton dissociation efficiency for different type of planar morphologies, depending on their thickness.[1] They investigated a perfectly mixed and uniform morphology (noted 'Mixed' below), a bilayer morphology ('PHJ' for planar heterojunction) with one of the layers 20nm thicker than the other, a three layer morphology with a 1:1 perfectly mixed 10nm layer between the donor and the acceptor layers ('PMHJ' for planar mixed heterojunction), and a three layer morphology with a pillar structured 10nm layer between the donor and the acceptor layers ('Chess' for chessboard). The reader is referred to Yang's paper for more details on the morphologies and material parameters. Figure 1 shows the side-by-side qualitative match between the results of Yang and the results of our model ($\overline{\eta_d}$: $the$ average of the local dissociation efficiencies,).

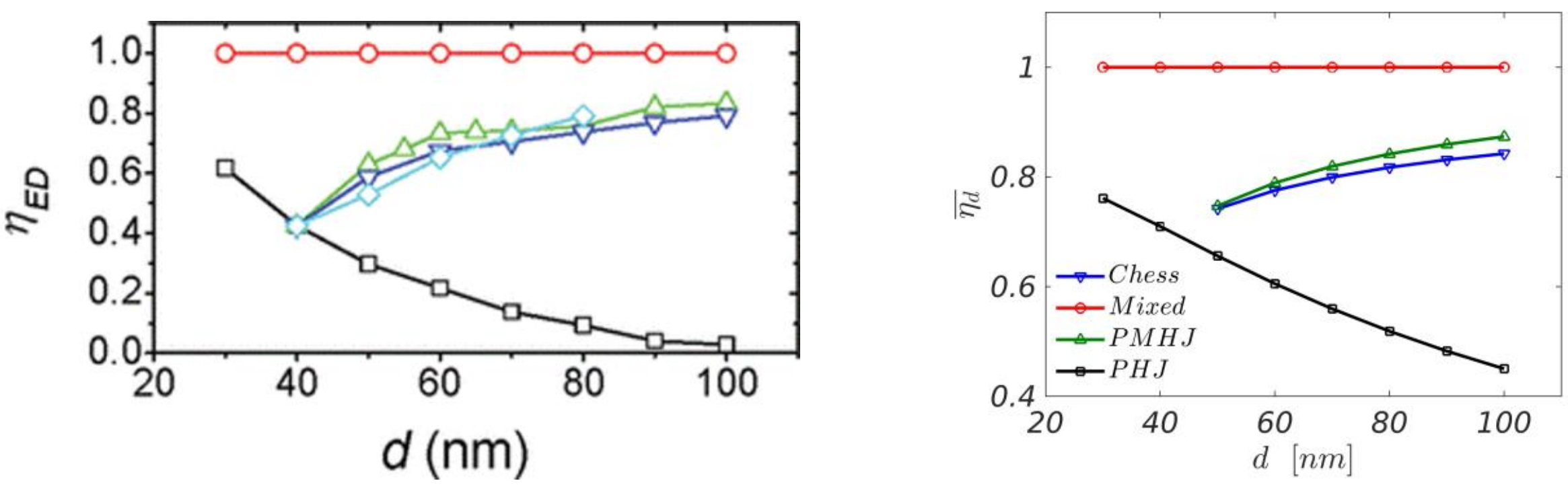

*Figure 1: Results of the exciton dissociation efficiency for different morphologies and for different thicknesses of the bulk heterojunction obtained by Yang and coworkers [1] (left) and our graph based model (right). The red, green, blue and black lines correspond to homogeneously mixed (mixed), planar-mixed (PMHJ), chessboard (Chess) and planar heterojunctions (PHJ), respectively. The intermediate layer thickness between a thin donor layer of 10 $nm$ and a thin acceptor layer of 30 $nm$ is varied for the planar-mixed and chessboard heterojunctions. In the planar heterojunction the acceptor layer is 20 nm thicker than the donor one for all thicknesses. Reprinted with permission from [1].*

Regarding the effect of thickness, the exciton dissociation efficiency is 1 for a homogenously mixed morphology. The exciton dissociation efficiency of a planar heterojunction decreases with increasing active layer thickness, because of the increase of the average distance of an exciton to the donor/acceptor interface. Finally, the exciton dissociation efficiency increases when the thickness of the intermediate layer is increased for the planar mixed and chessboard morphologies, because dissociation is ensured in the intermediate layer of increasing thickness: it is ensured with ideal efficiency for the mixed layer, while for the pillar structure the pillar size the efficiency is small enough for the excitons to reach the donor-acceptor interfaces. Overall, all results we report are consistent with those of Yang and coworkers. Note that we calculate the exciton dissociation efficiency into *collectable* free charge carriers whereas Yang et al. calculated the exciton dissociation efficiency. However, with the considered morphologies, all generated charge carriers are collectable, so that the calculated values are directly comparable.

### 1.1.2. Exciton dissociation and recombination for amorphous demixed morphologies of binary blends

Kodali investigated the effect of the donor-acceptor interface of a binary mixture undergoing spinodal decomposition on exciton dissociation and non-geminate recombination using 2D drift-diffusion modelling.[2] The investigated BHJ are two-phase amorphous co-continuous morphologies. Upon coarsening, the domain size increases and the interface area decreases. The reader is referred to Kodali's paper for more details on the morphologies and material parameters.

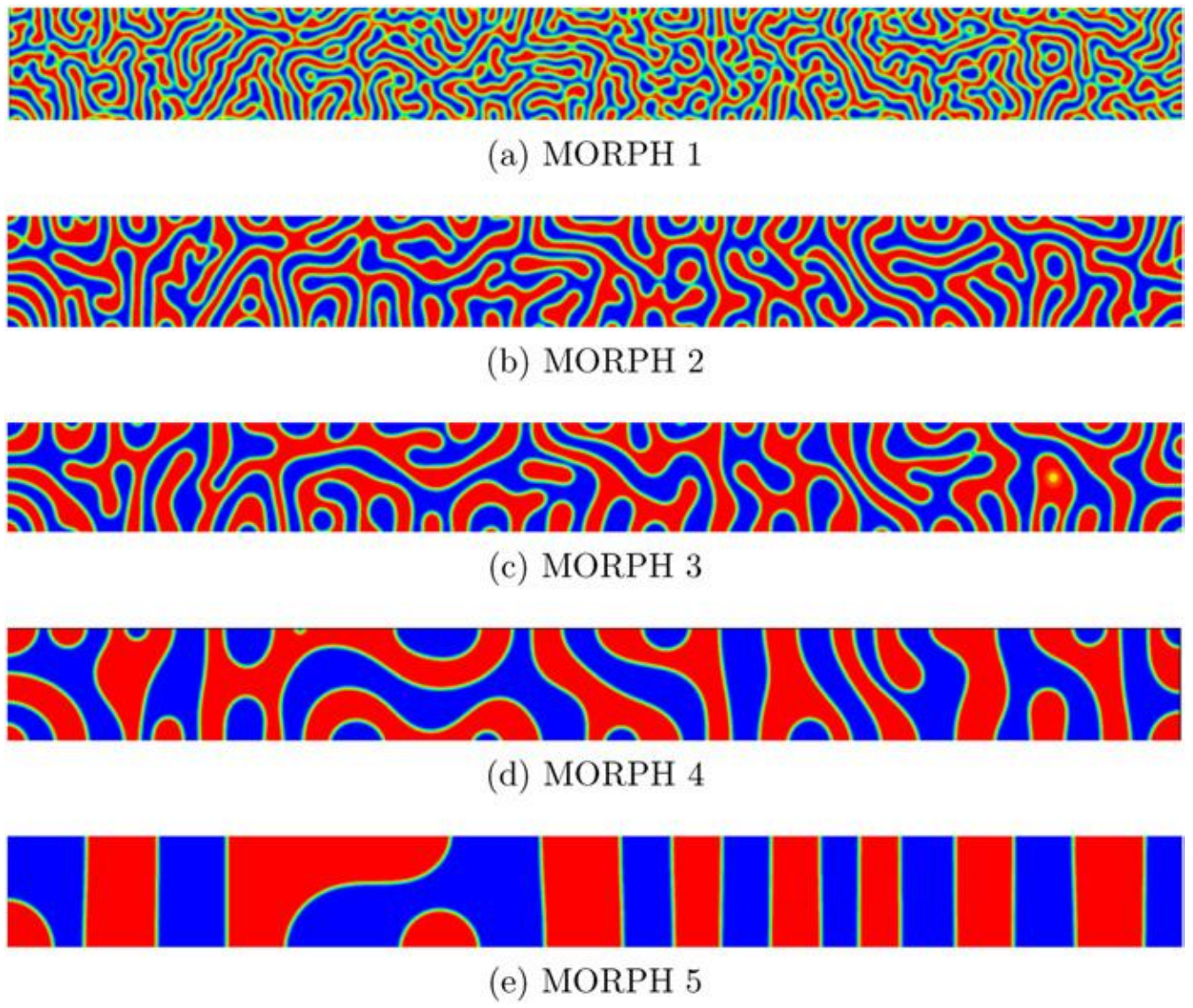

*Figure 2: Co-continuous morphologies investigated by Kodali et al. resulting from spinodal decomposition of an immiscible 1:1 amorphous donor-acceptor blend. Reprinted with permission from [3].*

In line with the results obtained by Kodali and coworkers, the dissociation efficiency descriptor calculated using our graph-based model increases with increasing interface area (Figure 3). Additionally, the non-geminate recombination morphology-aware descriptor increases with increasing interface area and follows a linear trend, as obtained in the results from Kodali and coworkers. Overall, our model reproduces a typical result, which is the increase of dissociation and recombination with interface. This is thought to be one of the fundamental reasons for the compromise between dissociation and recombination in BHJ. Note that in the work of Kodali, dissociation and recombination are assumed to happen at the donor-acceptor interfaces, and the values are integrated throughout the whole donor/acceptor interface available in the morphology. This differs from the descriptors' calculation method used in our work, whereby only the interfaces belonging to the CETP, e.g. where collectable free charge carrier can be generated, are considered (see main text). Despite this slight difference in the descriptor definition, our results compare very well with Kodali's results. In our results, the drop in exciton dissociation efficiency (into collectable free charge carrier) at large interface area originates in the large amount of isolated domains.

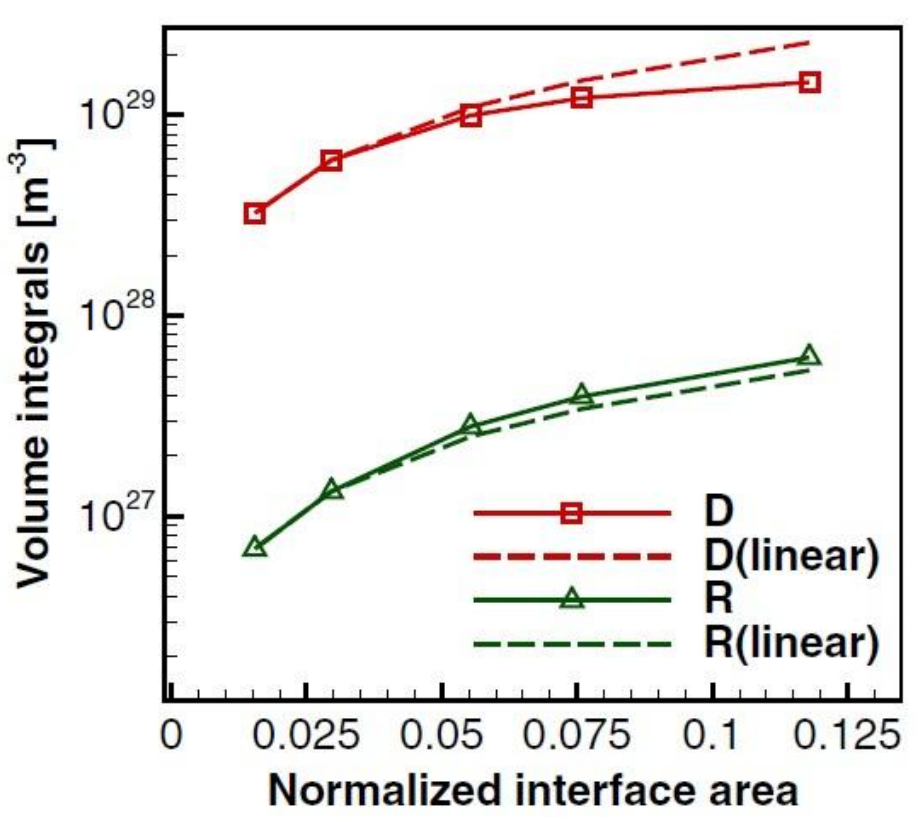

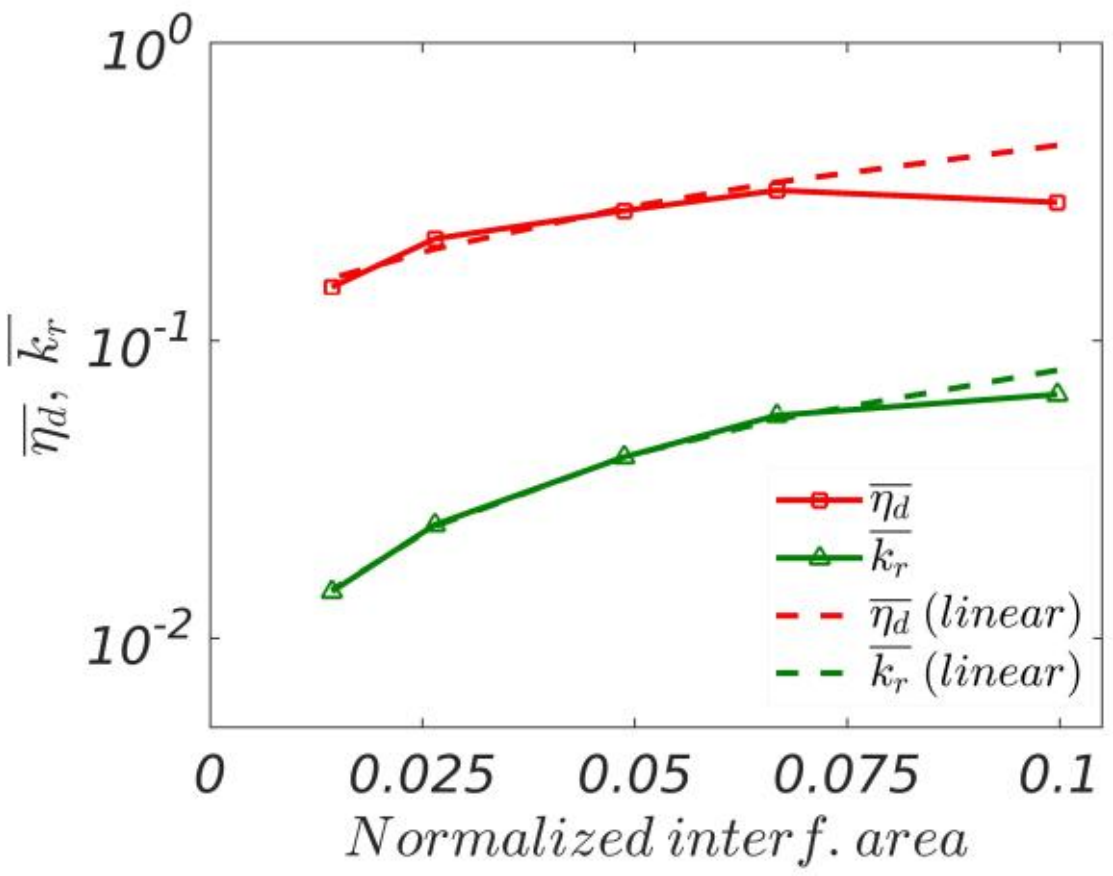

*Figure 3: Results of the exciton dissociation efficiency and non-geminate recombination depending on the normalized interface area for a morphology going through spinodal demixing obtained by Kodali and coworkers (left, reprinted with permission from [2] [3]) and our graph-based model (right). The simulated exciton dissociation efficiency and recombination descriptors are plotted with a solid line, and the dashed lines correspond to linear fits.*

Groves and coworkers investigated the effect of domain size and domain purity on the exciton dissociation in a binary mixture undergoing spinodal decomposition using Monte Carlo simulations.[4] Unfortunately, we were not able to reproduce exactly the morphologies used by Groves, but tried to generate similar BHJs. Nevertheless, we obtained very similar trends compared to those of Groves and coworkers (Figure 4). The exciton dissociation efficiency increases with decreasing domain purity, because the dissociation is favored in impure domains. As expected, and in-line with Kodali's findings, the exciton dissociation efficiency decreases with increasing domain size. For low purity phases, the minority material concentration is sufficient in both demixed donor and acceptor phase to allow for dissociation everywhere, and the dissociation efficiency is equal to 1.

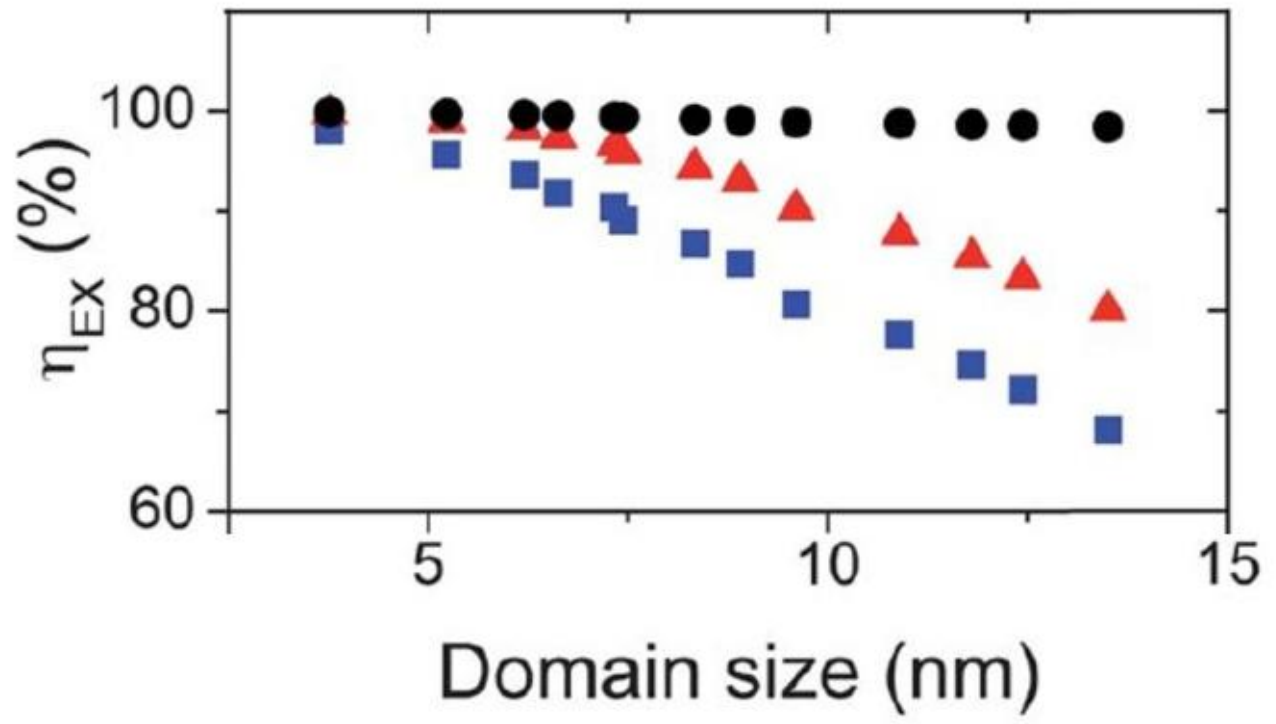

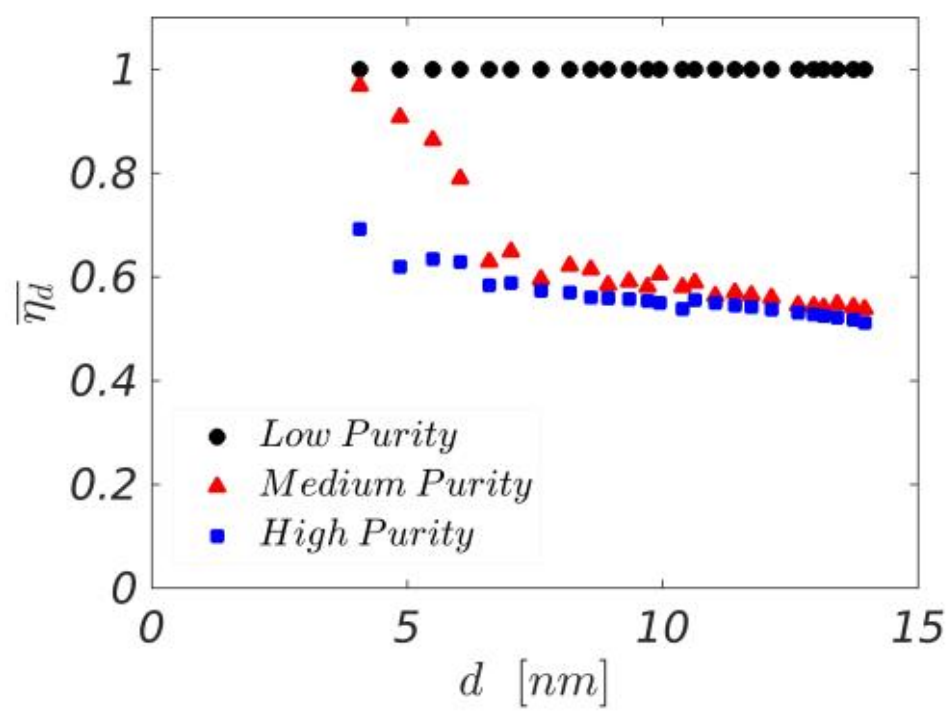

*Figure 4: Results of the exciton dissociation efficiency of a morphology undergoing spinodal demixing and for different purities of the spinodal domains obtained by Groves and coworkers[4] (left, Reprinted with permission from [4]) and our graph-based descriptor (right). The black, red and blue curves correspond to the results for very impure, impure and pure domains.*

### 1.1.3. Mobility for semi-crystalline morphologies of a single component

Geng and coworkers investigated the effect of crystallinity and grain density of a pure acceptor material using 3D master equation simulations.[5] Thereby, they varied the overall crystallinity and the grain density or, equivalently, the grain size. They assumed the mobility to be significantly larger in the crystalline domains and introduced an energy penalty for an electron hopping from the crystalline to the amorphous phase. On top of that, the hopping distance allowed for electron hopping across grain boundaries.

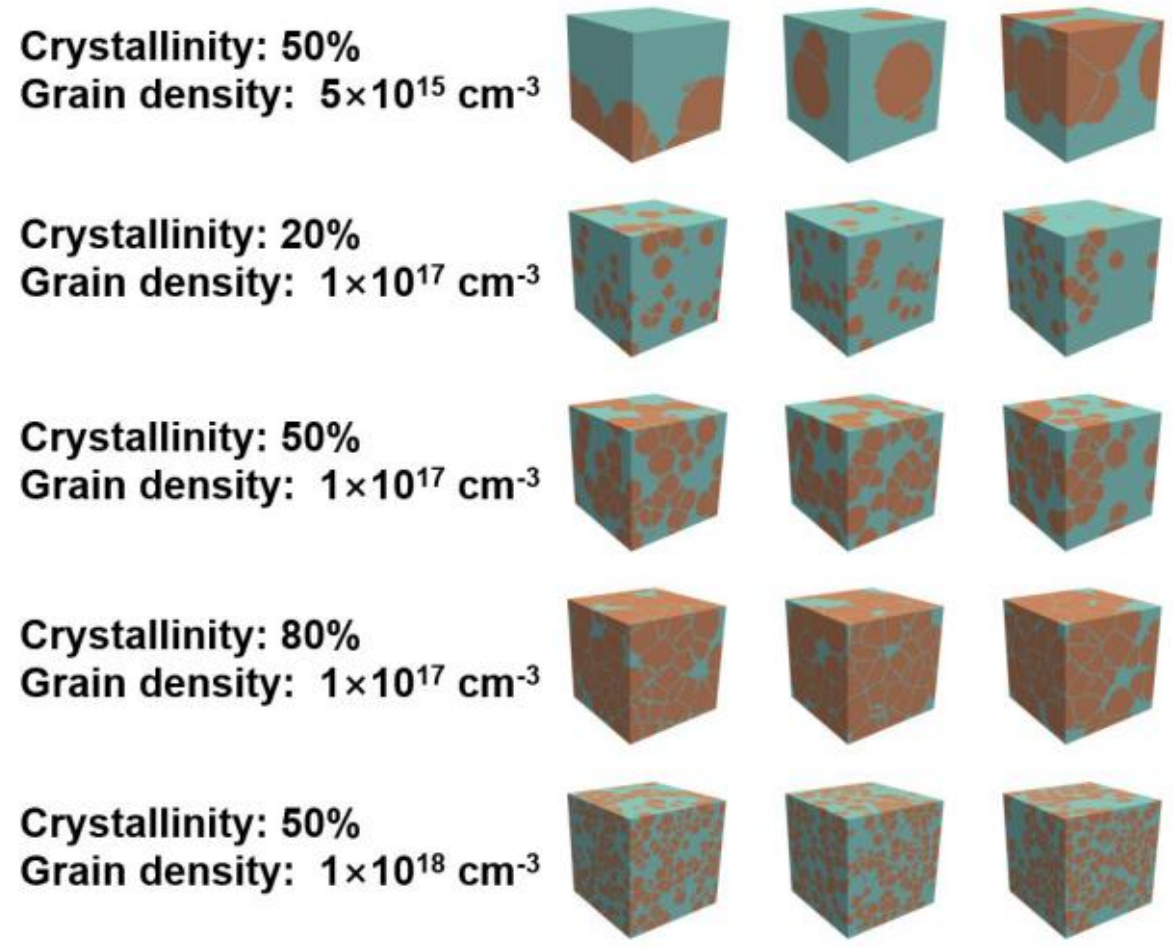

*Figure 5: Examples of 3D morphologies investigated by Geng, for various crystallinity and grain density values. Reprinted with permission from [5].*

In order to investigate the effect of crystallinity and grain size on mobility, in this work, we generated 2D morphologies with similar grain size and volume fractions. Assuming a sufficiently large penalty for crystal-amorphous hopping ($p_e \ll 1$, see main text) and using morphologies for which direct grain-grain contact is allowed and thus fast transport from one grain to another is allowed, the results of Geng et al. are qualitatively recovered (Figure 6): first, as expected, the mobility increases non-linearly with increasing crystallinity at fixed grain density. Second, the mobility decreases with increasing grain density. Third, at low crystallinity values, the mobility for a semi-crystalline morphology is lower than for a purely amorphous morphology.

To illustrate the origin of these three effects, we performed simulations with slightly different parameters. The non-linear increase of mobility comes from the harmonic averaging of mobility along streamlines (see main text, Equations 13-14), as can be seen from Figure 7 (left). The effect of grain density only appears with an energetic barrier for transport from the crystalline to the amorphous phase (see Figure 7, right). The higher the grain density, the larger the interface density, the larger the effect. The mobility at low crystallinity is lower than the mobility of the amorphous phase if the mobility at the crystal-amorphous interface is sufficiently low: interfaces then behave as blocking layers for charge transport. This effect is very strong up to very large crystallinities because the interface density increases with crystallinity (see Figure 7, right). It is fast crystal-crystal transport across grain boundaries that allows mobilities to increase significantly at large crystallinities (see Figure 6). This analysis is in-line with the discussion provided by Geng, which shows that our model not only allows to qualitatively recover Geng's results, but also that the underlying mechanisms are the proper ones.

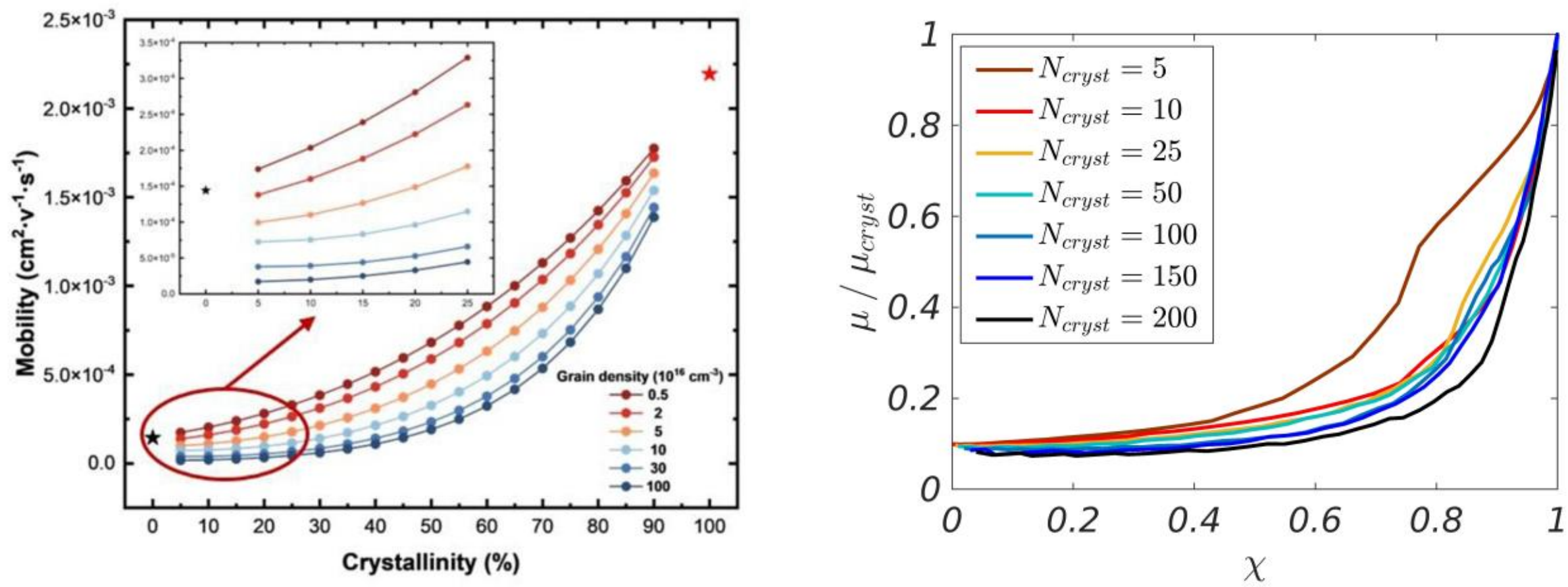

*Figure 6: Results on the mobility of a pure acceptor morphology depending on the crystallinity and grain density obtained by Geng and coworkers (left, reprinted with permission from [5]) and our descriptor from graph-based model (right). In our simulations, a penalty for crystal-amorphous hopping $p_e = 10^{-2}$ is used, and direct grain- grain contact is allowed. The number of crystals $N_{cryst}$ correspond to a given grain density / grain size.*

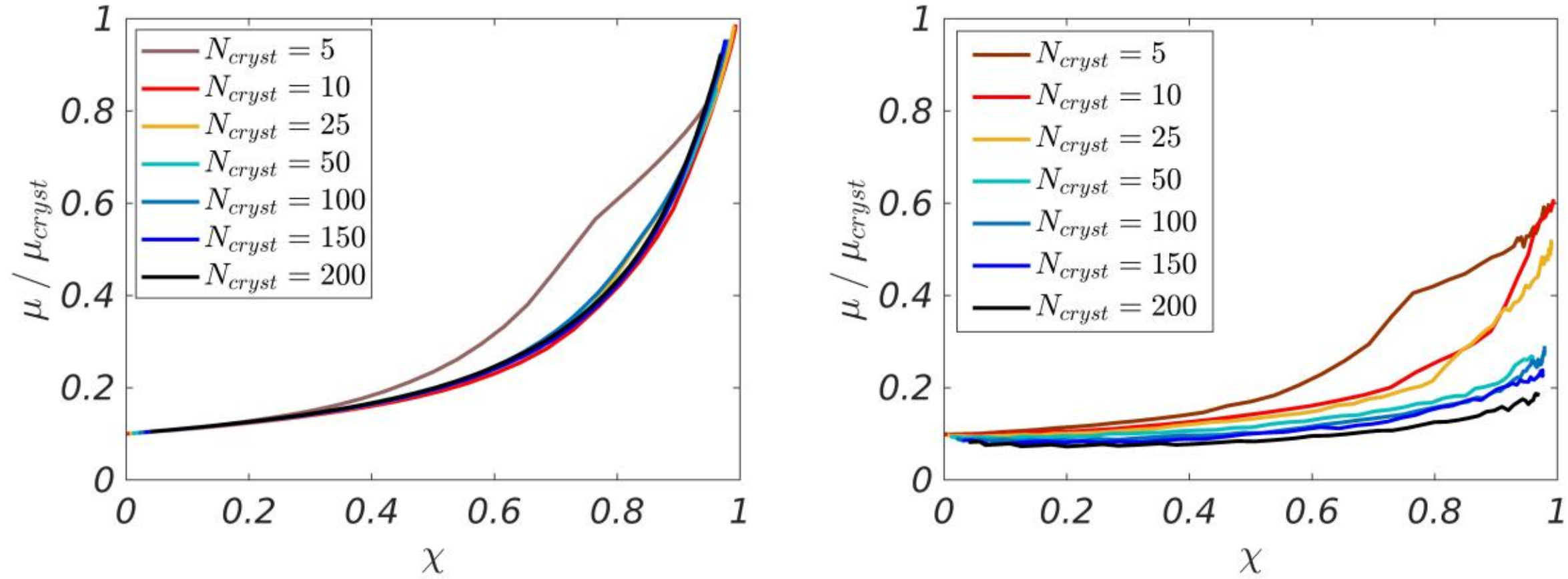

*Figure 7: Results on the mobility of a pure acceptor morphology depending on the crystallinity and grain density. (Left) Without any penalty for crystal-amorphous hopping. The non-linear mobility increase is present, but there is no impact of grain density, and the minimal mobility is equal to the mobility of the amorphous phase. (Right) With a penalty $p_e = 10^{-2}$ for crystal-amorphous hopping and without grain- grain contact. The crystal-amorphous hopping penalty is responsible for the effect of grain density and the mobilities below the amorphous mobility at low crystallinities (compare Figure 7, left). Forbidding fast transport across grain boundaries prevents mobilities converge quickly to the value of the crystalline phase at large crystallinities (compare Figure 6, left).*

## 1.2. Investigation of morphologies of increasing complexity

In addition to the comparison to advanced simulations results by other groups discussed above, we calculated the exciton dissociation, recombination and mobility descriptors on morphologies with varying complexity. The objective was to check if the model's behavior corresponds to physics-based intuition. To this aim, we generated different types of morphologies:

- A fully *homogeneous*, amorphous 1:1 donor acceptor morphology (morphology noted "HA" for homogeneous amorphous)
- Morphologies with a *lateral composition gradient* so that a thin donor phase is present on the left, a thin acceptor phase on the right, and a broad mixed phase in-between. The donor and acceptor phases are either amorphous (morphology noted "LG" for lateral gradient) or crystalline ("LGcr" for crystalline)
- *Bilayer* morphologies, the composition of both layers being homogeneous. Both layers can be either amorphous or crystalline. For the first morphology ("BL1") the thickness is bigger than for the others bilayer morphology, and both layers are amorphous. For the other bilayers ("BL2", "BL2crD", "BL2crA", BL2cr") respectively: both layers are amorphous, only the donor layer is crystalline, both layers are amorphous, only the acceptor layer is crystalline, both layers are crystalline.
- *Trilayer* morphologies with a donor phase layer, a mixed phase layer and an acceptor phase layer. All layers are amorphous. For the first trilayer ("TL1"), all layers are homogeneous, whereas a vertical composition gradient is present for the second trilayer ("TL2")
- Morphologies featuring *pillar-like structures* of separated donor and acceptor phases. The first one features thin (as compared to the exciton diffusion length) pillars traversing from the top to the bottom of the morphology, whereby both donor and acceptor phases are either amorphous or crystalline ("PI1", "PI1cr"). In the second one, the pillars are sandwiched between a donor and an acceptor layer ("PI2", "PI2cr"). In the third one, the pillars are not straight, but rather have a multi-sinusoidal, fractal structure ("PI3", "PI3cr").
- *Co-continuous*, two-phases amorphous morphologies resulting from spinodal decomposition at early stages ("COC1", small domain sizes, poor purity of demixed phases) and at late stages ("COC2", large domain sizes, high purity of demixed phases).
- Morphologies featuring *inclusions* (droplets and/or crystals) in a *matrix*. The two first ones ("IM1", "IM2") result from amorphous phase separation through nucleation and growth of a binary amorphous blend with 30%-70% vol. donor-acceptor blend ratio at early (small droplets) and late stages (large droplets), respectively. The next ones ("IM3", "IM4", "IM5") are three-phases morphologies featuring small elliptical donor and acceptor crystals in an amorphous mixed matrix, an amorphous donor matrix, and amorphous acceptor matrix, respectively. The two last ones are five phases morphologies featuring small elliptical donor and acceptor crystals in an amorphous demixed matrix, whereby the amorphous donor phase is the majority phase for "IM6", and the acceptor phase the majority phase for "IM7".

Figure 8 shows the exciton dissociation efficiency, recombination and mobility descriptors for all these morphologies, using the parameters mentioned in Table 1 of the main text, except equal exciton diffusion lengths $L_a = L_d = 10nm$. Without going into a detailed analysis of these results, the conclusion is that the model behaves as expected.

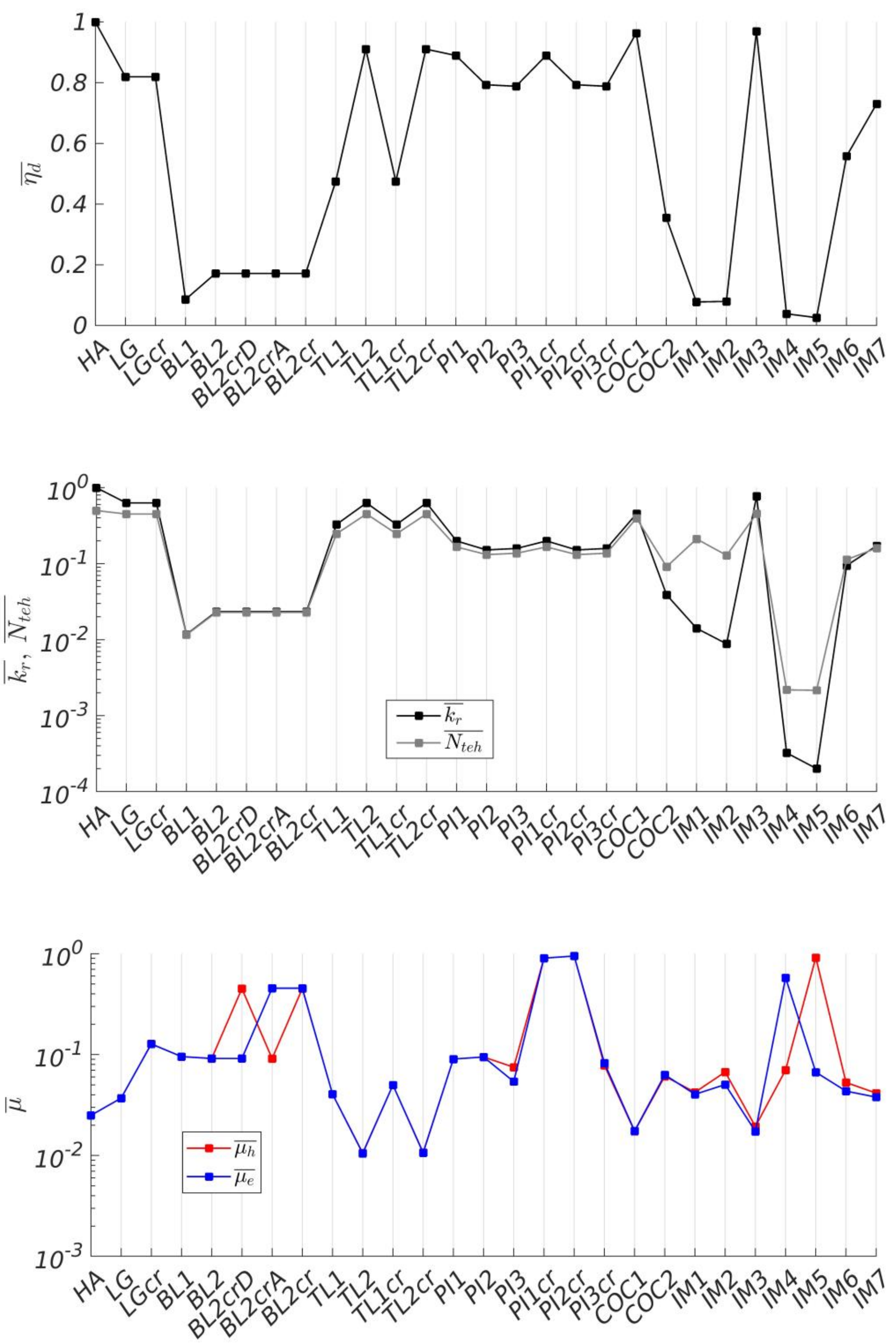

*Figure 8: Descriptors for the morphologies of varying complexity (from top): exciton dissociation efficiency, non-geminate recombination, and electrons and holes mobilities calculated.*

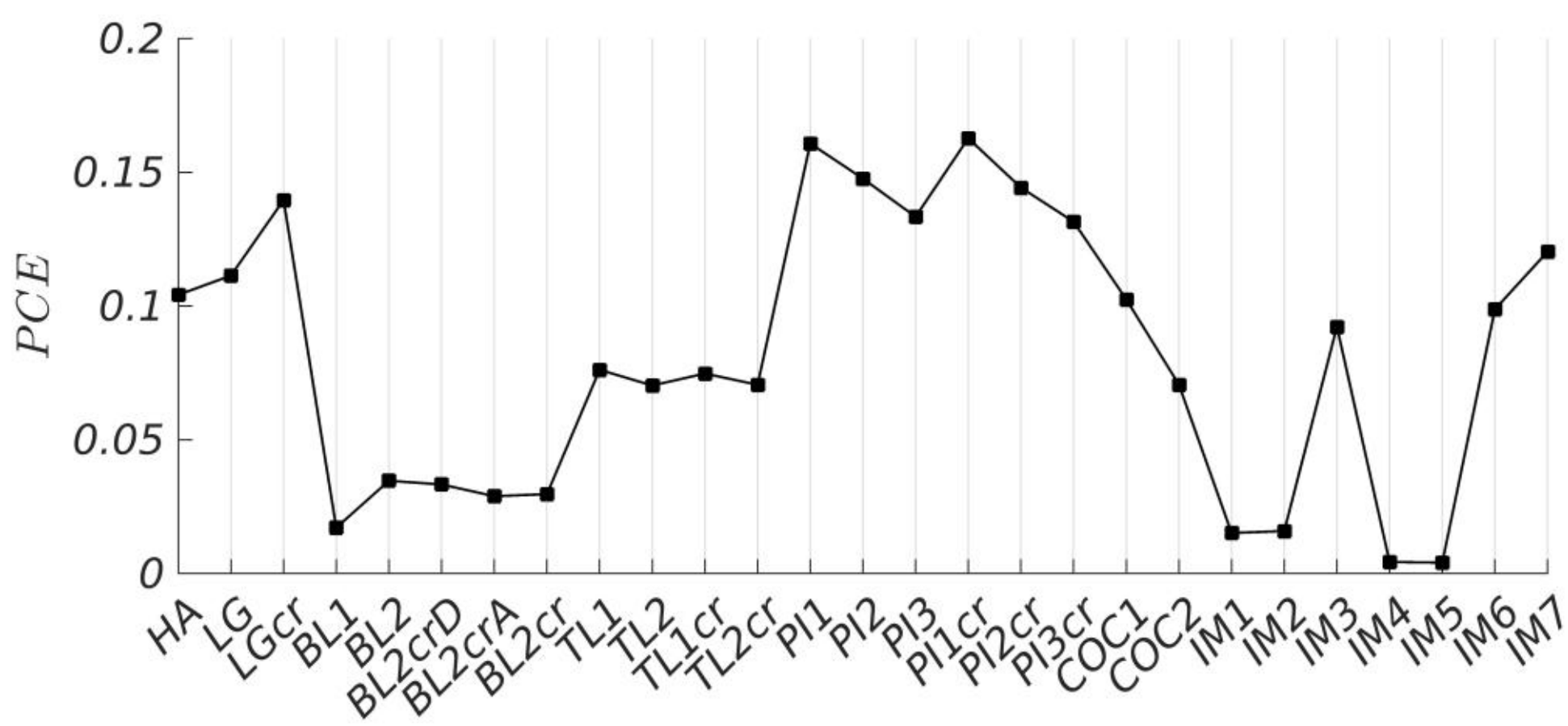

*Figure 9: PCE for the morphologies of varying complexity, assuming the same parameters for drift-diffusion modelling as compared to the main text (PM6-Y6 solar cell).*

## 2. Drift-diffusion simulation of a real PM6-Y6 solar cell; determination of ${\eta_d}^i$, ${\mu_h}^i$, ${\mu_e}^i$, ${k_r}^{wc}$ and ${N_{teh}}^{wc}$

The objective here is to get reasonable reference values for ${\eta_d}^i$, ${\mu_h}^i$, ${\mu_e}^i$, ${k_r}^{wc}$ and ${N_{teh}}^{wc}$, which reflect the optoelectronic properties of the considered donor-acceptor blend. For this, we choose to fit the JV curve of real PM6-Y6 solar cells in order to obtain dissociation efficiency, recombination and mobility values from the fit. Thereby, we base on the data of a stack chosen by Sharma and coworkers for solar cells with 16% efficiency,[6] because all data required for the drift-diffusion for the definition the stack are clearly reported, and because both JV curves in the dark and under one sun AM1.5 illumination are available.

The stack is composed of ITO / PEDOT:PSS / PM6:Y6 / P2G / Ag. Table 1 summarizes the model parameters used for the fitting of the dark and illuminated JV curves of PM6: Y6. For the transfer matrix formalism optical model, the ITO electrode is 180nm thick, the silver electrode 80nm thick, and the refractive indices of the PM6:Y6 absorber layer have been taken from ref. [7].

| *Layer properties* | *Symbol* | *Unit* | *PEDOT PSS* | *PM6 Y6* | *P2G* |
|---|---|---|---|---|---|
| *Thickness* [6] | $l$ | $nm$ | $20$ | $140$ | $5$ |
| *LUMO energy level* [6,8] | $E_{LUMO}$ | $eV$ | $-2.2$ | $-3.92$ | $-4.03$ |
| *HOMO energy level* [6,8] | $E_{HOMO}$ | $eV$ | $-5.23$ | $-5.18$ | $-6.05$ |
| *HOMO or LUMO density of states* | $N_v, N_c$ | $cm^{-3}$ | $5 \cdot 10^{20}$ | $5 \cdot 10^{20}$ | $5 \cdot 10^{20}$ |
| *Relative permittivity* [9–11] | $\varepsilon_r$ | - | $5$ | $3.74$ | $2.73$ |
| *Dissociation efficiency*[5] | $\eta_d$ | - | | $\mathbf{0.965}$ | |
| *Electron mobility* | $\mu_e$ | $cm^2V^{-1}s^{-1}$ | $10^{-7}$ | $\mathbf{10^{-3}}$ | $10^{-3}$ |
| *Hole mobility* | $\mu_h$ | $cm^2V^{-1}s^{-1}$ | $0.77$ | $\mathbf{3.2 \cdot 10^{-4}}$ | $10^{-7}$ |
| *Bimolecular recombination factor* | $k_r$ | $cm^3s^{-1}$ | | $\mathbf{6.4 \cdot 10^{-13}}$ | |
| *SRH trap density* | $N_t$ | $cm^{-3}$ | $0$ | $\mathbf{2.4 \cdot 10^{9}}$ | $0$ |
| *SRH trap energy level* | $E_{tr}$ | $eV$ | | $-4.55$ | |
| *SRH electron and hole capture coefficient* | $C_n, C_p$ | $cm^3s^{-1}$ | | $10^{-7}$ | |
| *Electron and hole surface recombination velocity* | $S_n, S_p$ | $cm\ s^{-1}$ | $\infty$ | | |
| *Cathode work function* [6] | $WF_c$ | $eV$ | $-3.92$ | | |
| *Anode work function* [6] | $WF_a$ | $eV$ | $-4.95$ | | |
| *Series resistance* | $R_s$ | $\Omega cm^{-2}$ | $\mathbf{0.78}$ | | |
| *Shunt resistance* | $R_{sh}$ | $\Omega cm^{-2}$ | $\mathbf{6.3 \cdot 10^{3}}$ | | |

*Table 1: Drift-diffusion parameters used for the fitting of the experimental JV curves of the PM6:Y6 organic solar cell measured by Sharma and coworkers.[6] The fixed parameters values are in black and the fitted parameter values in red.*

The experimental and fitted JV curves are shown in Figure 10. The values obtained for the fitted dissociation efficiency, mobilities and bimolecular recombination are in line with reported literature data. [7] [12] [13] [14] [15] [16]

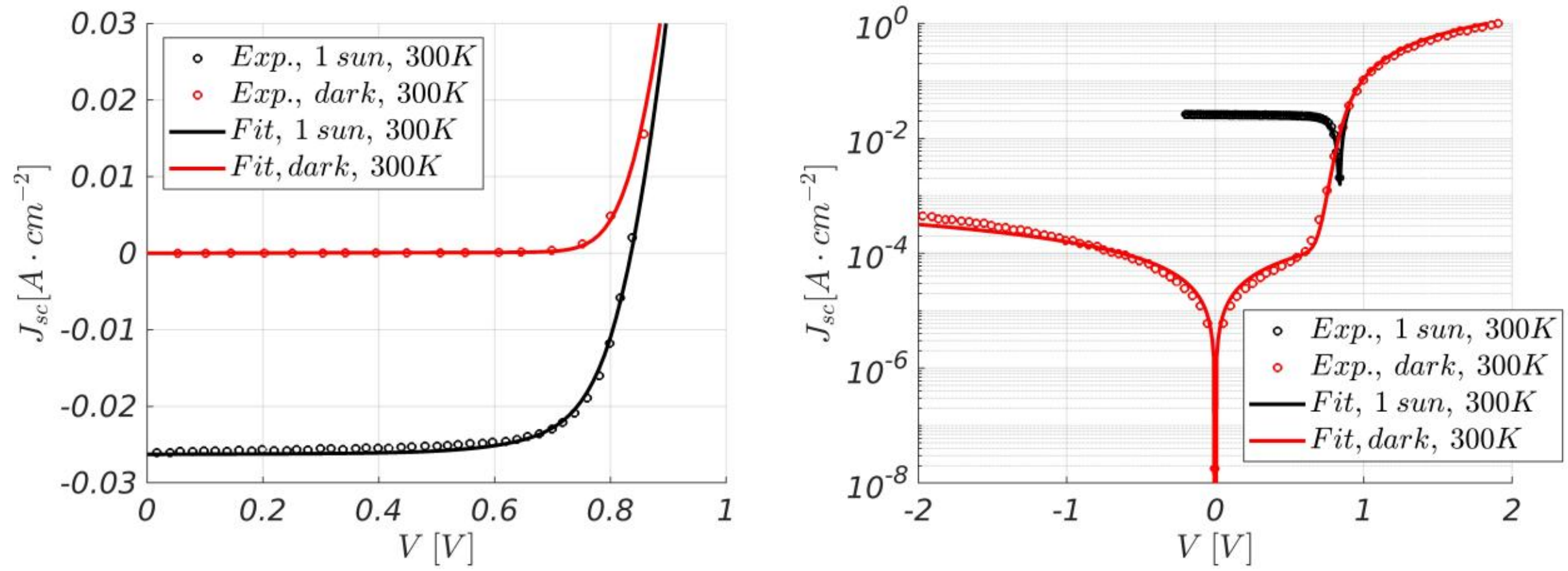

*Figure 10: experimental and fitted JV curves in the dark and under one sun illumination of the PM6: Y6 solar cell from ref. [6]. (left) Linear scale for the y-axis (right) Same data with a logarithmic scale for the y-axis.*

These values are used as a reference point for the drift diffusion simulations performed in the main text. We choose ${\eta_d}^i = 1$. For simplicity, we consider the donor and acceptor materials to have similar ideal mobilities if fully crystalline, about one decade above the fitted values, e.g. ${\mu_e}^i = {\mu_h}^i = 10^{-2}\ cm^2V^{-1}s^{-1}$. Similarly, the "worst case" bimolecular recombination factor is set to ${k_r}^{wc} = 6.4 \cdot 10^{-12}\ cm^3s^{-1}$ and the "worst case" trap density to ${N_{teh}}^{wc} = 2.4 \cdot 10^{10}\ cm^{-3}$.

Statistics on the morphology-dependent values over the time evolution of the six morphologies discussed in the main text show that $\overline{\eta_d}$ varies in the whole range $[0\ 1]$, the mobilities $\overline{\mu_h}$ and $\overline{\mu_e}$and the bimolecular recombination prefactor $\overline{k_r}$ in the range $[10^{-3}\ 1]$ (see Figure 11). Our (somewhat arbitrary at this stage) choice ensures that the values for the real cell belong to these variation ranges, with the idea that the real cell is rather very good but in principle still not optimal, except for exciton dissociation efficiencies. Note that within these ranges, the JV curves are very sensitive to mobility and exciton dissociation variations, and insensitive to the trap density variations, e.g. the trap density is anyway low enough that it does not impact the photovoltaic performance. The JV curves are partly sensitive to the bimolecular recombination factor: the performance obviously increases with decreasing recombination factor, until an asymptote is reached, whereby bimolecular recombination becomes negligible and does not play a role anymore. This asymptote is reached for the lowest ${k_r}^{DD}$ values.

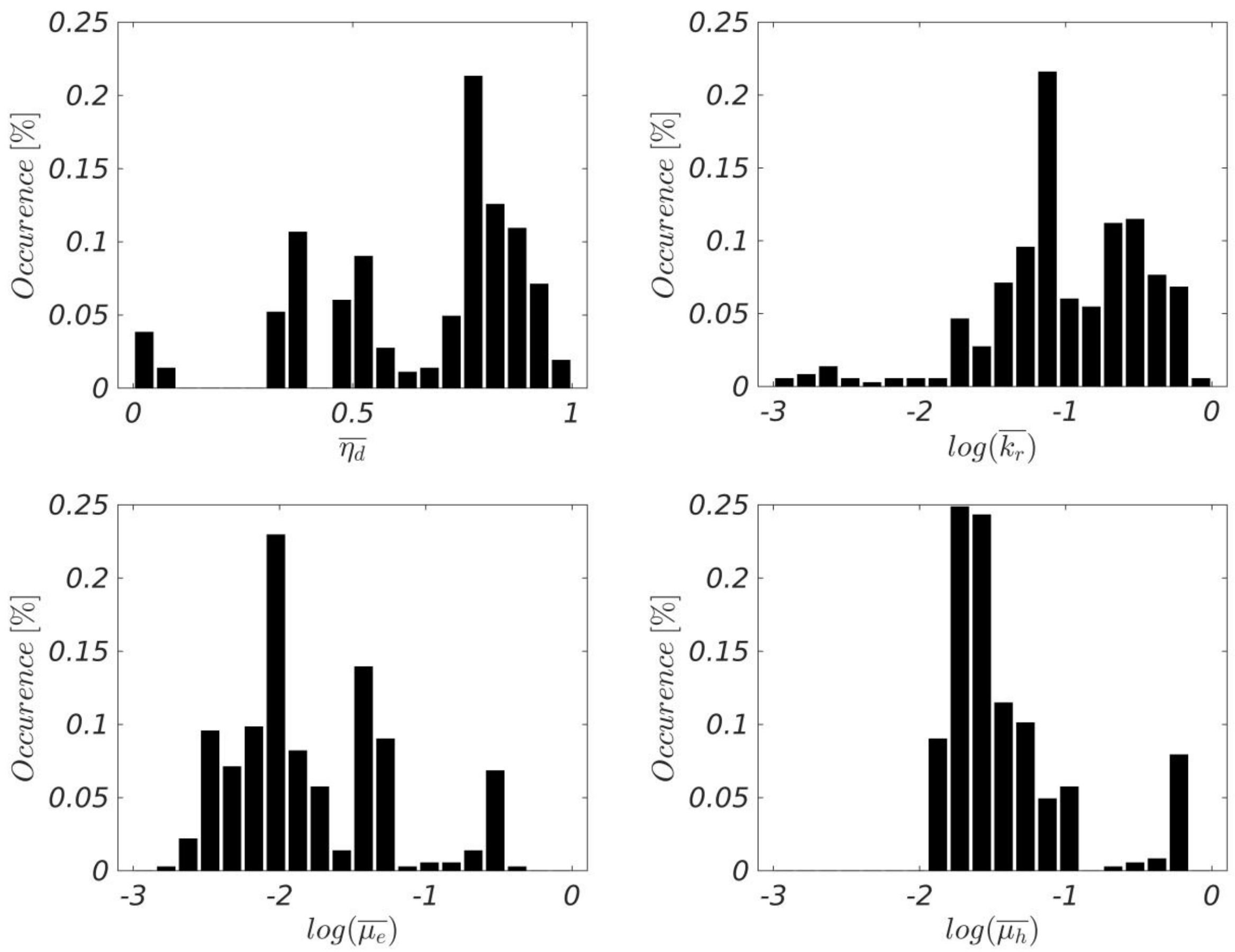

*Figure 11: Morphology-aware descriptors probability distributions for the six morphologies M1 to M6 and for all ageing times. (Top left) Exciton dissociation efficiency. (Top right) Non-geminate recombination. (Bottom Left) Electrons mobility (Bottom Right) Holes mobility.*

# 3. Detailed data to the structure-property investigations for morphologies M1 to M6

## 3.1. Morphology and electronic properties of the as cast films

The following table show morphological parameters describing the as-cast films and their corresponding performance results. We introduce in this section the following additional parameters $\phi_{Cr,d}$, $\phi_{Cr,a}$, $\phi_{Am,d}$, $\phi_{Am,d}$, and $\phi_{Am,M}$ which correspond to the overall volume fraction of the crystalline donor, crystalline acceptor, amorphous donor, amorphous acceptor and amorphous mixed phases in the BHJ morphology.

| | M1 | M2 | M3 | M4 | M5 | M6 |
|---|---|---|---|---|---|---|
| $\boldsymbol{\phi_d}$ | 0.6 | 0.6 | 0.6 | 0.6 | 0.6 | 0.5 |
| $\boldsymbol{\phi_a}$ | 0.4 | 0.4 | 0.4 | 0.4 | 0.4 | 0.5 |
| $\boldsymbol{\chi_d}$ | 0.037 | 0.530 | 0.150 | 0.179 | 0.419 | 0.190 |
| $\boldsymbol{\chi_a}$ | 0.000 | 0.016 | 0.315 | 0.000 | 0.110 | 0.193 |
| $\boldsymbol{\phi_{Cr,d}}$ | 0.024 | 0.324 | 0.094 | 0.110 | 0.265 | 0.103 |
| $\boldsymbol{\phi_{Cr,a}}$ | 0.000 | 0.007 | 0.129 | 0.000 | 0.045 | 0.111 |
| $\boldsymbol{\phi_{Am,d}}$ | 0.021 | 0.043 | 0.094 | 0.462 | 0.036 | 0.000 |
| $\boldsymbol{\phi_{Am,a}}$ | 0.122 | 0.215 | 0.102 | 0.338 | 0.103 | 0.000 |
| $\boldsymbol{\phi_{Am,M}}$ | 0.834 | 0.412 | 0.581 | 0.089 | 0.552 | 0.785 |
| $\boldsymbol{\overline{\eta_d}}$ | 0.958 | 0.891 | 0.907 | 0.366 | 0.905 | 0.982 |
| $\boldsymbol{\overline{\mu_e}}$ | 0.0088 | 0.0059 | 0.0058 | 0.0567 | 0.0102 | 0.0151 |
| $\boldsymbol{\overline{\mu_h}}$ | 0.0254 | 0.0127 | 0.0204 | 0.0309 | 0.0123 | 0.0162 |
| $\boldsymbol{\overline{k_r}}$ | 0.6725 | 0.3198 | 0.3968 | 0.0438 | 0.4376 | 0.7333 |
| $\boldsymbol{\overline{G_{ex}}}$ | $1.29 \cdot 10^{28}$ | $1.33 \cdot 10^{28}$ | $1.27 \cdot 10^{28}$ | $1.30 \cdot 10^{28}$ | $1.31 \cdot 10^{28}$ | $1.32 \cdot 10^{28}$ |
| $\boldsymbol{G}$ | $1.23 \cdot 10^{28}$ | $1.18 \cdot 10^{28}$ | $1.15 \cdot 10^{28}$ | $4.77 \cdot 10^{27}$ | $1.19 \cdot 10^{28}$ | $1.29 \cdot 10^{28}$ |
| $\boldsymbol{J_{sc}}$ | -0.0254 | -0.0233 | -0.0228 | -0.0107 | -0.0246 | -0.0271 |
| $\boldsymbol{V_{oc}\ [V]}$ | 0.779 | 0.797 | 0.790 | 0.832 | 0.791 | 0.781 |
| $\boldsymbol{FF}$ | 0.37 | 0.36 | 0.35 | 0.78 | 0.40 | 0.41 |
| $\boldsymbol{PCE}$ | 0.074 | 0.067 | 0.064 | 0.070 | 0.078 | 0.088 |

## 3.2. Green equation and simulated fill factor for a simplified drift-diffusion modelling setup

As can be seen in Figure 8 of the main text, the simulated fill factor follow qualitatively the trend of the empiric Green equation proposed by Neher,[17] but there is no quantitative match. This has to do with the fact that the drift diffusion simulation setup used for the main text is quite sophisticated, including three layers, series resistances, misaligned energy levels... whereas Neher performed one-layer simulations without all these features. In order to mimic Neher's approach, we also calculated the JV curves of the six morphologies upon ageing with a simple one-layer drift-diffusion setup, without energy level misalignment at the cathode and anode, traps, shunt and series resistances. Under these conditions, a quantitative agreement with Neher's equation is fairly recovered (Figure 12).

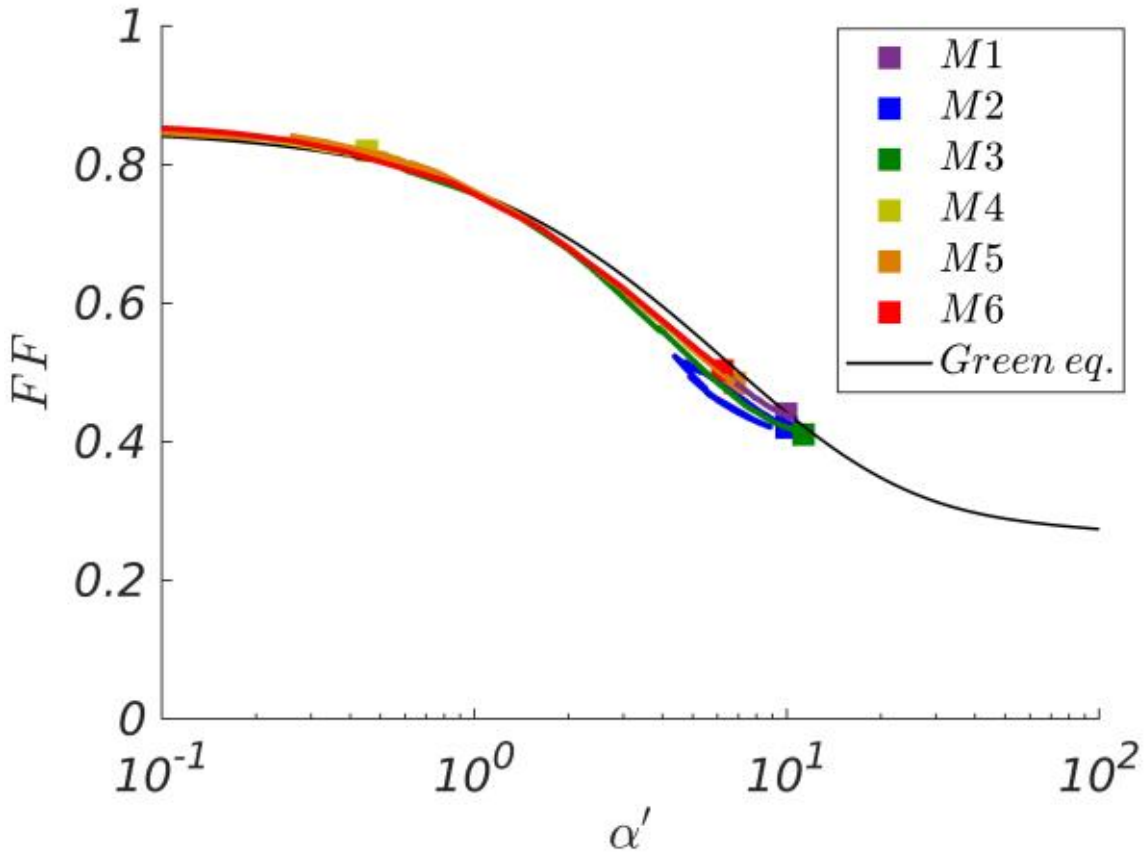

*Figure 12: FF plotted against $\alpha'$ (Equation 24 of the main text) and Green equation adapted by Neher (Equation 23 of the main text), using a one-layer simple drift-diffusion model similar to Neher's work.*

## 3.3. Evolution of the absorption spectra under thermal loading

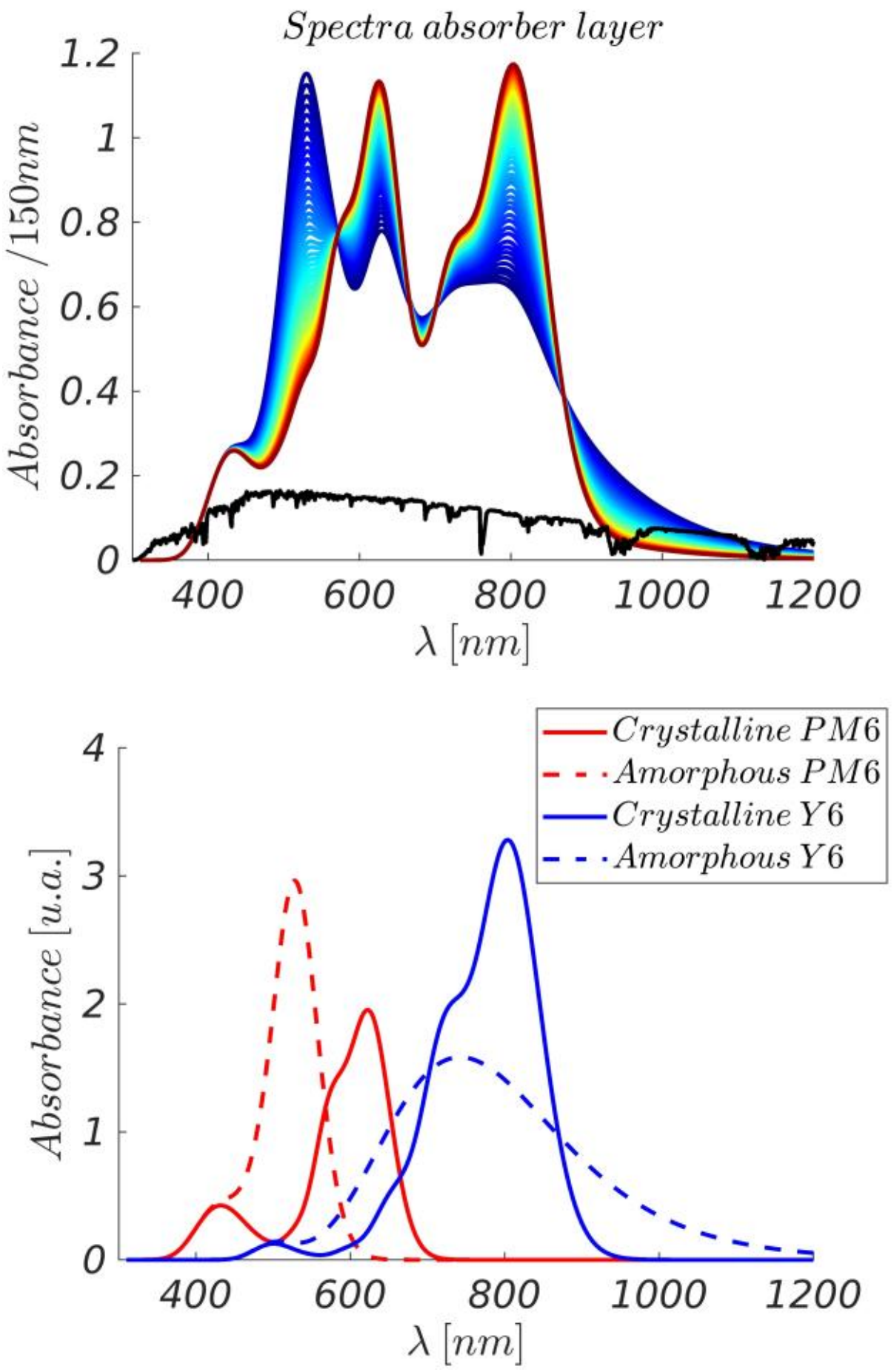

*Figure 13: (Top) Evolution of the absorption spectrum of the BHJ morphology M5 under thermal loading, due to both PM6 and Y6 crystallization. The colors of the curves encode the time, from blue (beginning of thermal evolution) to red (end of thermal evolution). For comparison, the black curve shows the AM1.5 spectrum. (Bottom) As a guide to the eye, the spectra of the crystalline and amorphous PM6 and Y6, as shown in the Figure 2 of the main text.*

## 3.4. Drop of exciton dissociation efficiency upon thermal evolution of morphology M5

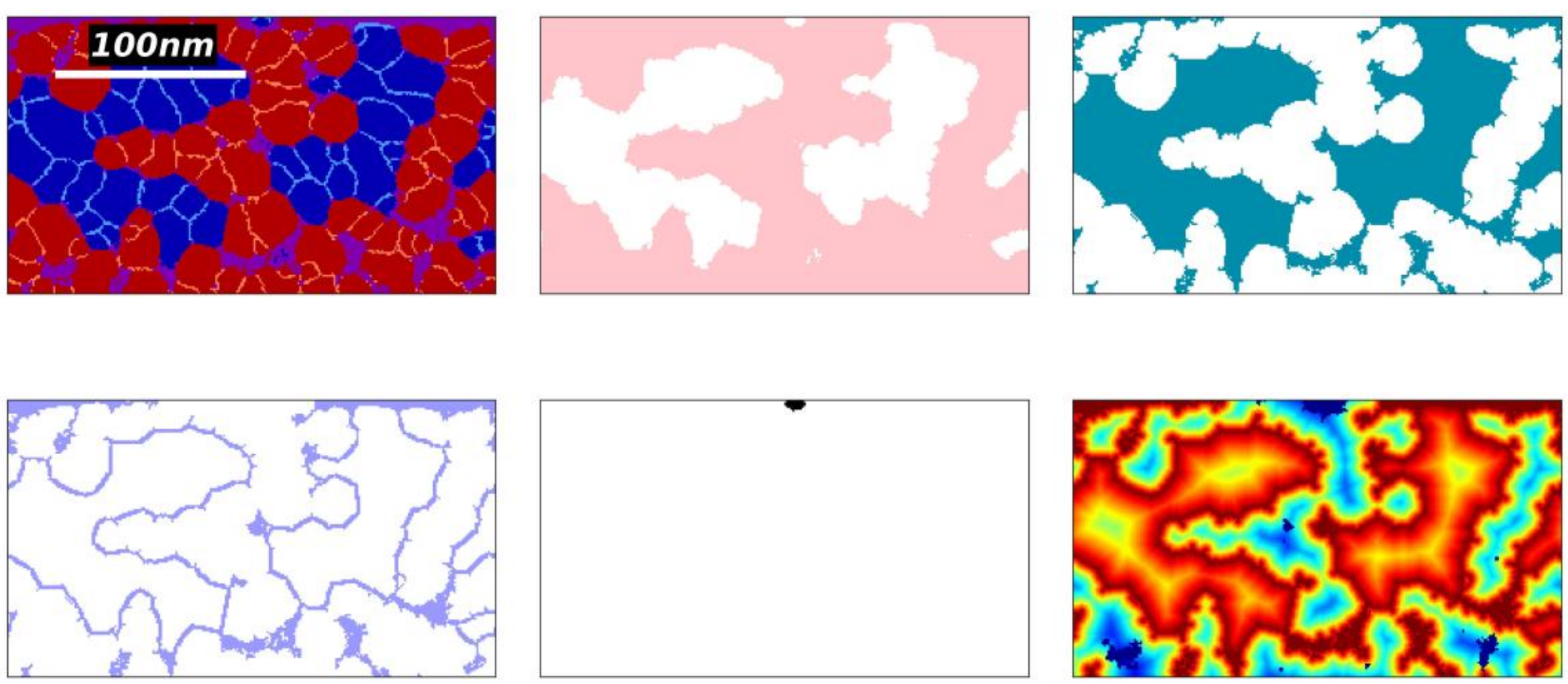

*Figure 14: Functional regions and exciton dissociation efficiency for morphology M5 just* ***before*** *the exciton dissociation efficiency drop. From left to right and top to bottom: phase type (donor amorphous in light red, donor crystalline in dark red, acceptor amorphous in light blue, acceptor crystalline in dark blue, mixed amorphous in purple), effective hole transport phase region (EHTP, salmon), effective electron transport phase region (EETP, turquoise), region common to both effective transport phase (CETP, lilac), isolated domains (black), exciton dissociation efficiency into collectable free charge carriers - the values scale linearly from 0 (dark blue) to 1(dark red). The larger part of the CETP is connected to the cathode (bottom of the BHJ) by only two veins, at the right end of the BHJ and at about one third of its width.*

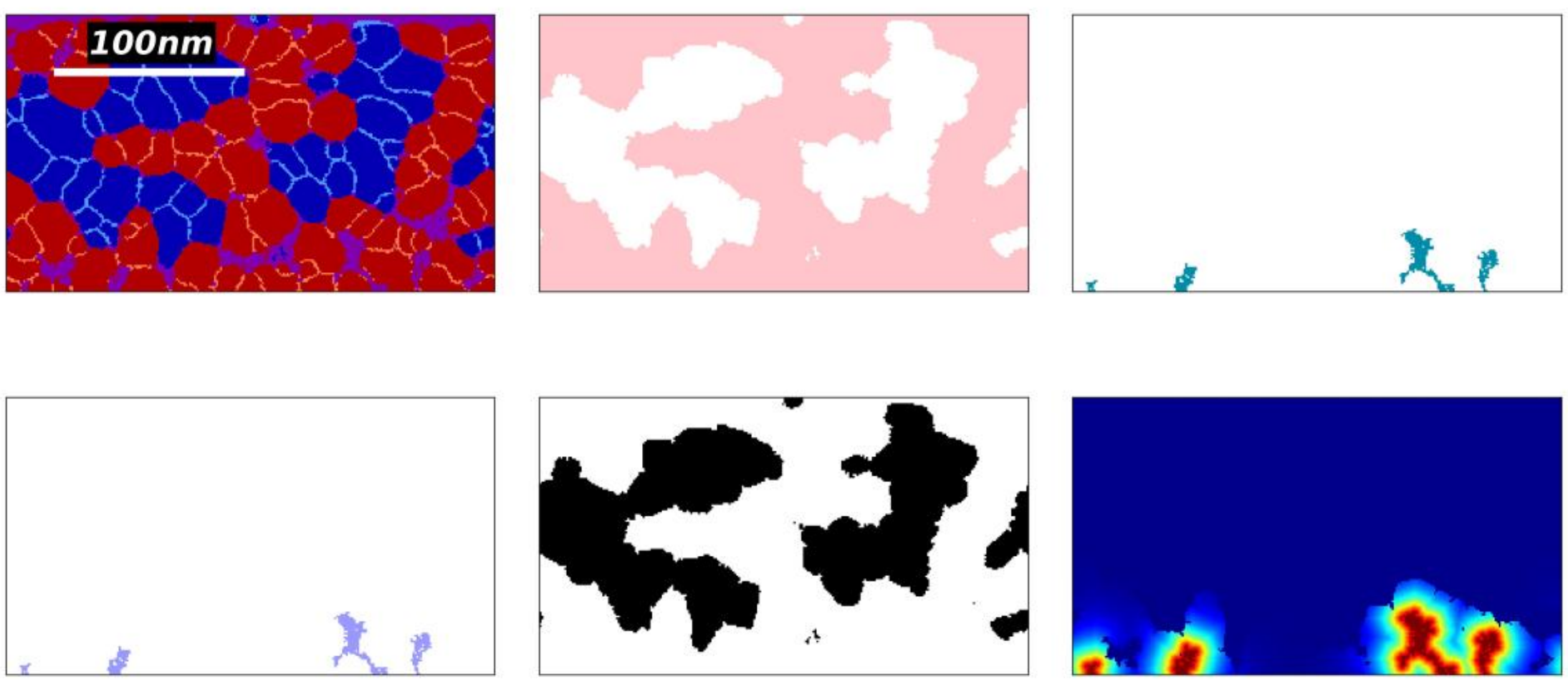

*Figure 15: Functional regions and exciton dissociation efficiency for morphology M5 just* ***after*** *the exciton dissociation efficiency drop. Panels show the same information as in Figure 15. The two veins discussed above have been cut, reducing the CETP to small domains close to the cathode.*

## 3.5. Relationship between bimolecular recombination rate and volume fraction of CETP

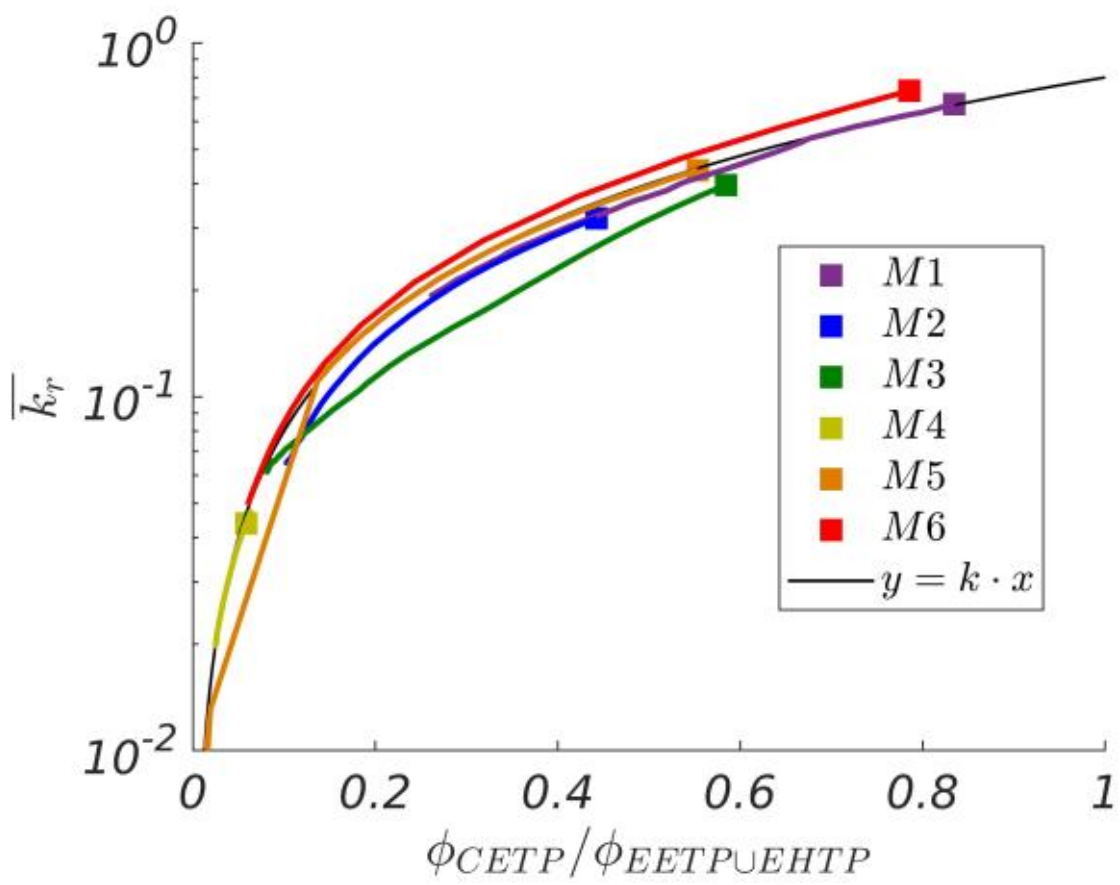

*Figure 16: Average bimolecular recombination rate for morphologies M1 to M6 during thermal evolution, depending on the ratio between the volume fraction of CETP and the volume fraction of effective transport phases $\phi_{CETP}/\phi_{EETP \cup EHTP}$. Each curve represents the time evolution for a given morphology, whereby the large symbols mark the as-cast morphology.*

# 4. Performance evolution under thermal loading for morphologies M1 to M6 calculated with various model parameters

## 4.1. PM6:Y6 with different model parameters

In the main text, it is assumed that bimolecular recombination is non-Langevin. A strong penalty has also been used for crystal to amorphous phase hopping. The figure below shows the sensitivity of the final PCE results to both of these assumptions.

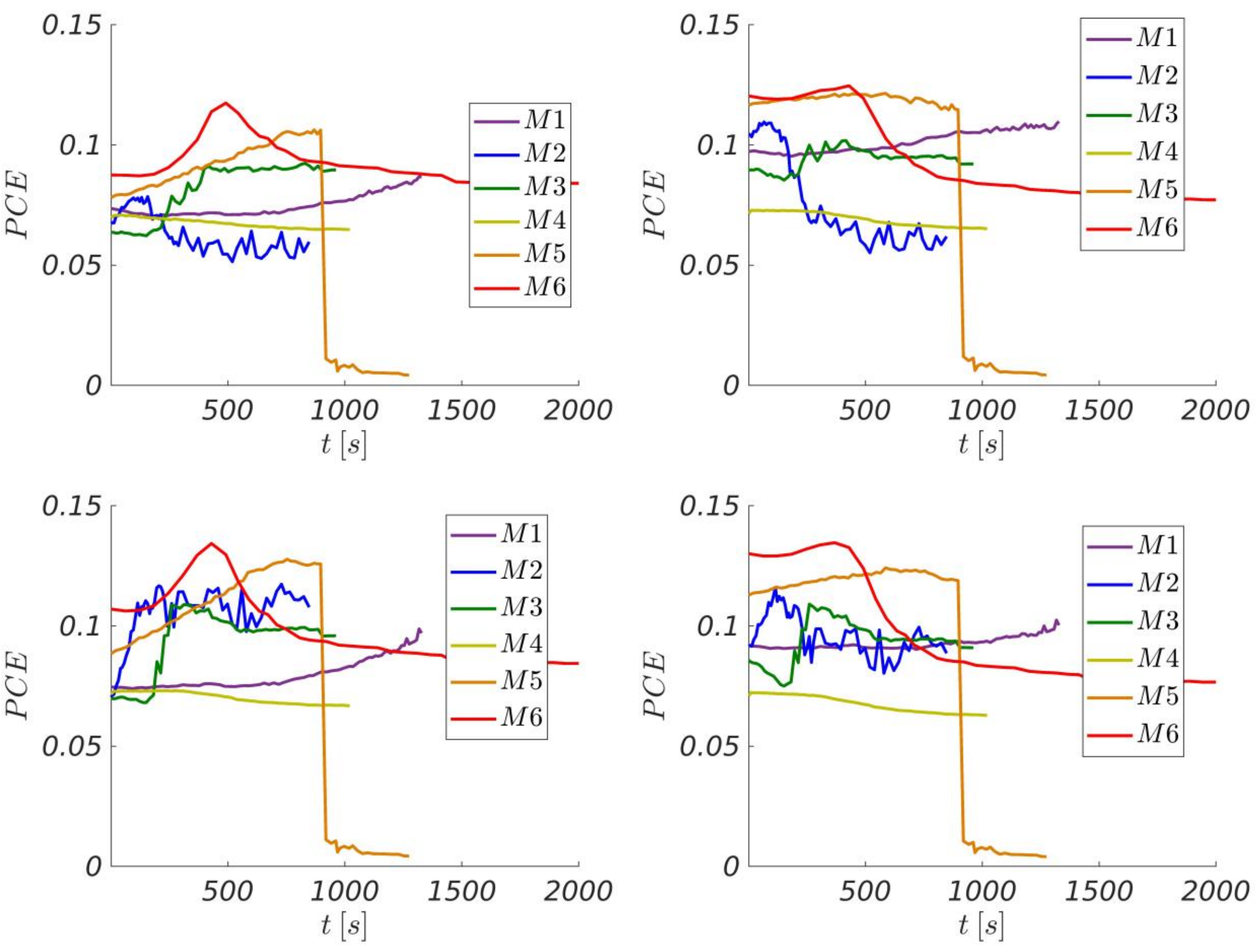

*Figure 17: Time dependent PCE for morphologies M1 to M6. (Top left) Non-Langevin recombination and crystal-amorphous hopping penalty $p_e = p_h = 0.01$ (reproduction of Figure 7 of the main text for reference). (Top right) Langevin recombination and crystal-amorphous hopping penalty $p_e = p_h = 0.01$ (Bottom left) Non-Langevin recombination and no crystal-amorphous hopping penalty (Bottom right) Langevin recombination and no crystal-amorphous hopping penalty.*

## 4.2. P3HT: PCBM materials parameters

In this section, we analyze the exact same morphologies M1 to M6 considered in the main text, but now assuming that they correspond to morphologies of P3HT: PCBM absorber layers. This allows us to evaluate the sensitivity of the approach to the optoelectronic parameters of the pure materials, which are indeed input parameters of the method. This also allows us to illustrate the interplay between morphological properties and material properties in the structure-property relationship.

For the calibration of the reference values ${\eta_d}^i$, ${\mu_h}^i$, ${\mu_e}^i$, ${k_r}^{wc}$ and ${N_{teh}}^{wc}$, we used the same approach as for PM6: Y6. We base our drift diffusion simulations on the data of a stack chosen by Alam and coworkers.[18] The stack is composed of ITO / PEDOT:PSS / P3HT: PCBM / Al. Table 2 summarizes the model parameters used for the fitting of the dark and illuminated JV curves of P3HT: PCBM (Figure 18). The ITO electrode is 180nm thick and the aluminum electrode 200nm thick.

| *Layer properties* | *Symbol* | *Unit* | *PEDOT PSS* | *P3HT PCBM* |
|---|---|---|---|---|
| *Thickness* [18,20,21] | $l$ | $nm$ | 107 | 146 |
| *LUMO energy level* [8,21] | $E_{LUMO}$ | $eV$ | $-2.2$ | $-4.1$ |
| *HOMO energy level* [8,21] | $E_{HOMO}$ | $eV$ | $-5.23$ | $-4.98$ |
| *HOMO or LUMO density of states* | $N_v, N_c$ | $cm^{-3}$ | $5 \cdot 10^{20}$ | $5 \cdot 10^{20}$ |
| *Relative permittivity* [11,22] | $\varepsilon_r$ | - | 5 | 3.4 |
| *Dissociation efficiency* | $\eta_d$ | - | | $\mathbf{0.92}$ |
| *Electron mobility* | $\mu_e$ | $cm^2V^{-1}s^{-1}$ | 10 | $\mathbf{1.3 \cdot 10^{-3}}$ |
| *Hole mobility* | $\mu_h$ | $cm^2V^{-1}s^{-1}$ | 10 | $\mathbf{1.6 \cdot 10^{-4}}$ |
| *Bimolecular recombination factor* | $k_r$ | $cm^3s^{-1}$ | | $\mathbf{1.2 \cdot 10^{-12}}$ |
| *SRH trap density* | $N_t$ | $cm^{-3}$ | 0 | $\mathbf{10^{10}}$ |
| *SRH trap energy level* | $E_{tr}$ | $eV$ | | $-4.54$ |
| *SRH electron and hole capture coefficient* | $C_n, C_p$ | $cm^3s^{-1}$ | | $10^{-7}$ |
| *Electron and hole surface recombination velocity* | $S_n, S_p$ | $cm\ s^{-1}$ | $\infty$ | |
| *Cathode work function*[19] | $WF_c$ | $eV$ | $-4.1$ | |
| *Anode work function*[19] | $WF_a$ | $eV$ | $-4.95$ | |
| *Series resistance* | $R_s$ | $\Omega cm^{-2}$ | $\mathbf{6}$ | |
| *Shunt resistance* | $R_{sh}$ | $\Omega cm^{-2}$ | $\mathbf{5.1 \cdot 10^4}$ | |

*Table 2: Drift-diffusion parameters used for the fitting of the experimental JV curves of the P3HT:PCBM organic solar cell measured by Alam and coworkers.[18] The fixed parameters values are in black and the fitted parameter values in red.*

The experimental and fitted JV curves are shown in Figure 18. The values obtained for the fitted dissociation efficiency, mobilities and bimolecular recombination are fully in line with reported literature data. [14] [16]

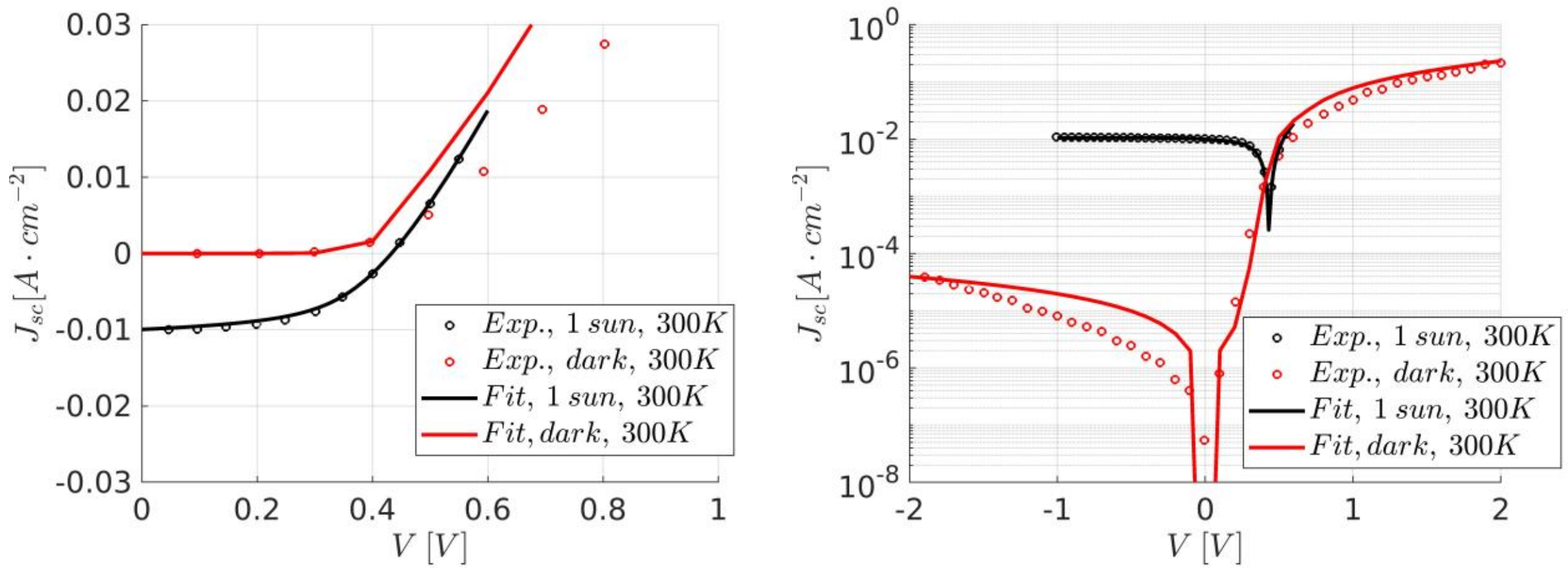

*Figure 18: Experimental and fitted JV curves in the dark and under one sun illumination of the P3HT: PCBM solar cell from ref. [18]. (left) linear scale for the y-axis (right) same data with a logarithmic scale for the y-axis.*

These values are used as a reference point for the drift diffusion simulations performed in the main text. We choose ${\eta_d}^i = 1$. We consider the donor and acceptor materials to have ideal mobilities if fully crystalline about one decade above the fitted values. ${\mu_e}^i = 10^{-2}\ cm^2V^{-1}s^{-1}$ and ${\mu_h}^i = 3.2 \cdot 10^{-3}\ cm^2V^{-1}s^{-1}$. Similarly, the "worst case" bimolecular recombination factor is set to ${k_r}^{wc} = 1.2 \cdot 10^{-11}\ cm^3s^{-1}$ and the "worst case" trap density to ${N_{teh}}^{wc} = 10^{11}\ cm^{-3}$. For optical modelling, it is assumed that only P3HT is able to absorb light, according to the spectra of the amorphous and the crystalline phases shown in Figure 19.

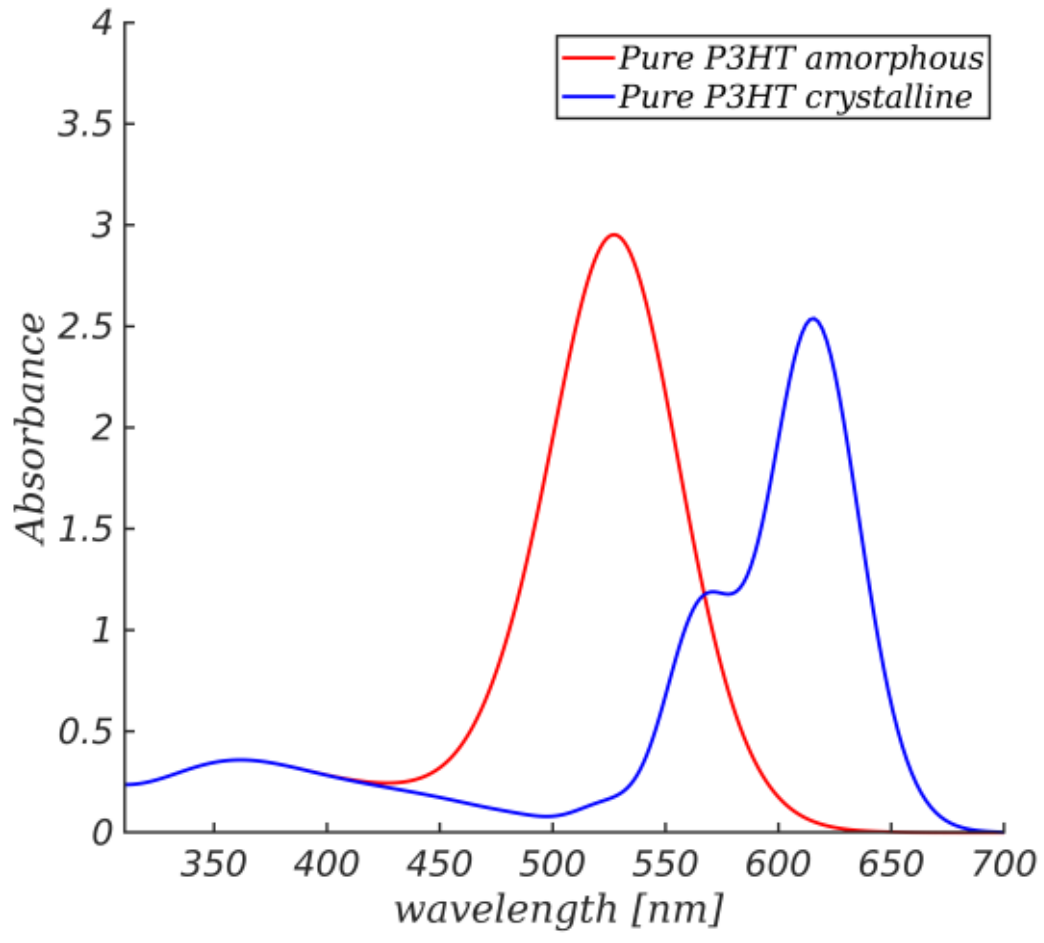

*Figure 19: Absorption spectrum of crystalline and amorphous P3HT. The absorption of PCBM is neglected.*

With this setup, the time-dependent JV curves can be calculated and the corresponding PCE extracted. Figure 20 below shows that the sensitivity of the photovoltaic performance is different compared to PM6:Y6, due to the different, material system-related balance between recombination, hole and electron mobility. This means that the sensitivity of OSC performance to BHJ morphology depends on the chosen material system.

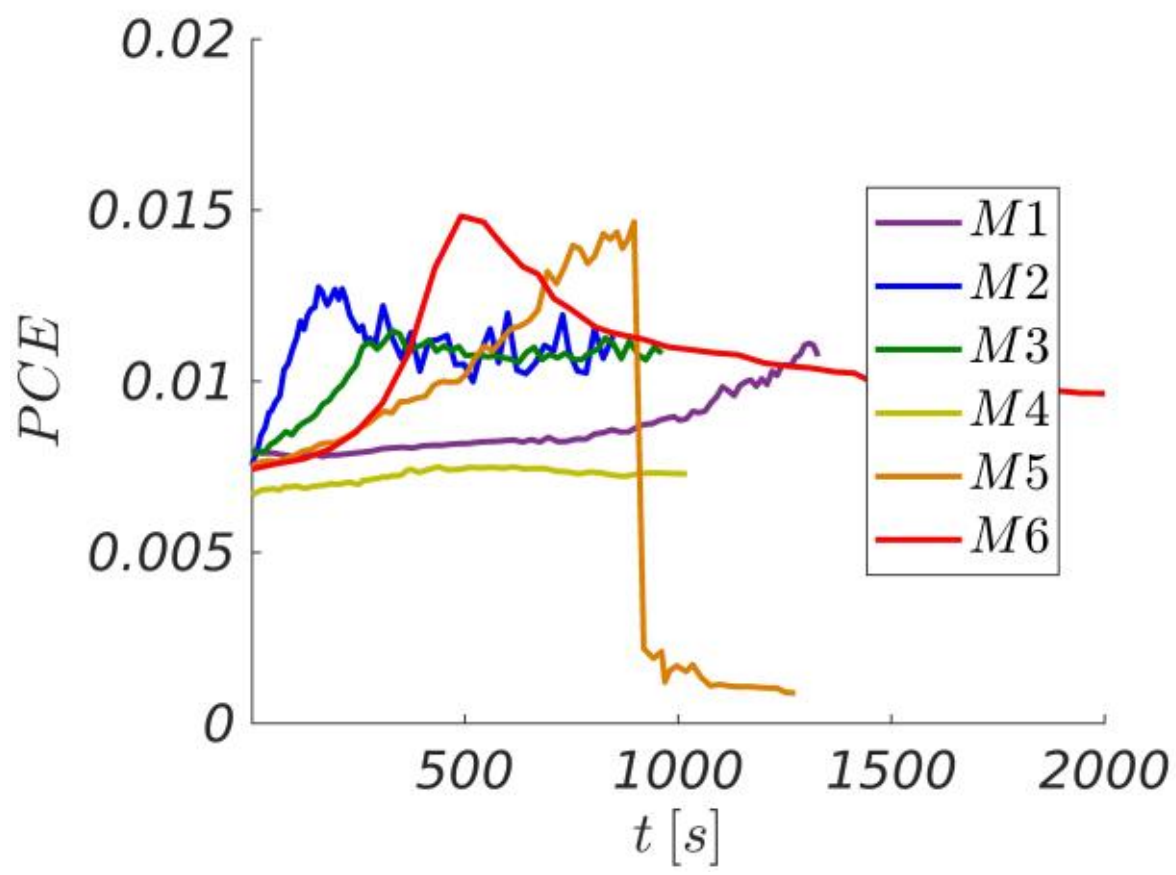

*Figure 20: Time dependent PCE for morphologies M1 to M6 using optoelectronic parameters suited for a P3HT: PCBM solar cell. Non-Langevin recombination and crystal-amorphous hopping penalty* $p_e = p_h = 0.01$*. The effect of P3HT:PCBM vs. PM6:Y6 parameters can be estimated by comparing with Figure 17, top left.*